\documentclass[aps, prb, 10pt, twocolumn, showpacs,
preprintnumbers, superscriptaddress, amsmath, amssymb]{revtex4-1}
\usepackage[pdftex]{graphicx}
\usepackage[ngerman,english]{babel}
\usepackage[ansinew]{inputenc}
\usepackage{natbib}
\usepackage{color}
\usepackage{siunitx}
\usepackage{calc}

\definecolor{green2}{rgb}{.0, .58, 0}

\begin{document}
\title{Dynamic cantilever magnetometry of individual CoFeB nanotubes}
\author{B. Gross}
\affiliation{Departement Physik, Universit\"{a}t Basel, 4056 Basel, Switzerland}
\author{D. P. Weber}
\affiliation{Departement Physik, Universit\"{a}t Basel, 4056 Basel, Switzerland}
\author{D. R{\"u}ffer}
\affiliation{Laboratoire des Mat\'eriaux Semiconducteur, Institut des Mat\'eriaux, Ecole Polytechnique F\'ed\'erale de Lausanne, 1015 Lausanne, Switzerland}
\author{A. Buchter}
\affiliation{Departement Physik, Universit\"{a}t Basel, 4056 Basel, Switzerland}
\author{F. Heimbach}
\affiliation{Lehrstuhl f\"ur Physik funktionaler Schichtsysteme, Physik Departement E10, Technische Universit\"at M\"unchen, 85747 Garching, Deutschland}
\author{A. Fontcuberta i Morral}
\affiliation{Laboratoire des Mat\'eriaux Semiconducteur, Institut des Mat\'eriaux, Ecole Polytechnique F\'ed\'erale de Lausanne, 1015 Lausanne, Switzerland}
\author{D. Grundler}
\affiliation{Laboratoire des Mat\'eriaux Magn\'etiques Nanostructur\'es and Magnoniques, Institut des Mat\'eriaux, Ecole Polytechnique F\'ed\'erale de Lausanne, 1015 Lausanne, Switzerland}
\author{M. Poggio}
\affiliation{Departement Physik, Universit\"{a}t Basel, 4056 Basel, Switzerland}
\date{\today}%
%
\begin{abstract}
  We investigate single CoFeB nanotubes with hexagonal cross-section
  using dynamic cantilever magnetometry (DCM).  We develop both an
  analytical model based on the Stoner-Wohlfarth approximation and a
  broadly applicable numerical framework for analyzing DCM
  measurements of magnetic nanostructures.  Magnetometry data show the
  presence of a uniformly magnetized configuration at high external
  fields with $\mu_0 M_s = 1.3 \pm \SI{0.1}{\tesla}$ and non-uniform
  configurations at low fields.  In this low-field regime, comparison
  between numerical simulations and DCM measurements supports the
  existence of flux-closure configurations.  Crucially, evidence of
  such configurations is only apparent because of the sensitivity of
  DCM to single nanotubes, whereas conventional measurements of
  ensembles are often obscured by sample-to-sample inhomogeneities in
  size, shape, and orientation.
  
%
%
\end{abstract}
\maketitle
%
%
\section{Introduction}
\label{sec:intro}

Applications ranging from dense magnetic memories
\cite{parkin_magnetic_2008}, to magnetic sensing
\cite{maqableh_low-resistivity_2012} and imaging
\cite{khizroev_direct_2002, poggio_force-detected_2010,
  campanella_nanomagnets_2011} have motivated the synthesis and study
of a wide range of nanometer-scale magnets.  At these size-scales,
geometry plays a crucial role in determining the magnetization
configurations that are stable.  Ferromagnetic nanotubes are a
particularly interesting morphology of nanomagnet, since the lack of a
magnetic core can make flux-closure magnetization configurations more
favorable than uniform single domain states.  These configurations minimize
magnetostatic energy and therefore produce minimal stray fields, e.g.\
reducing interactions between nanomagents in densely packed magnetic
memories.  A variety of stable configurations have been predicted at
low fields and at remanence, including a global vortex configuration,
where the spins point circumferentially around the tube, multi-domain
states composed of uniform and vortex domains, and an onion state,
consisting of two oppositely oriented circumferential domains plus two
uniform domains oriented in direction of the tube axis
\cite{castano_metastable_2003, wang_spin_2005, topp_internal_2008,
  landeros_equilibrium_2009, landeros_domain_2010,
  streubel_equilibrium_2012, ruffer_magnetic_2012,
  buchter_reversal_2013, streubel_magnetic_2014}. Flux-closure
configurations are particularly interesting since, during
magnetization reversal, they avoid the Bloch point structure and
thereby result in a fast and reproducible reversal process
\cite{landeros_equilibrium_2009}.  There are a variety of theories
describing the reversal process for such nanotubes
\cite{landeros_reversal_2007, escrig_crossover_2008,
  bachmann_size_2009, landeros_domain_2010,
  albrecht_experimental_2011, streubel_rolled-up_2013}, via a
propagating vortex, a transverse domain wall, or a mixed multi-domain
combination of the former.

Given their small magnetic moment, however, measurements of
magnetization and magnetization reversal in ferromagnetic nanotubes
have mostly been conducted on large ensembles
\cite{bachmann_ordered_2007, daub_ferromagnetic_2007,
  bachmann_size_2009, rudolph_ferromagnetic_2009,
  chong_multilayered_2010, albrecht_experimental_2011,
  escrig_crossover_2008}.  Difficultly in controlling the distribution
of size, shape, and orientation, as well as the interactions between
nanotubes complicate the interpretation of these results.  Here, we
avoid these complications by investigating individual CoFeB nanotubes
by dynamic cantilever magnetometry (DCM) \cite{rossel_active_1996,
  harris_integrated_1999, stipe_magnetic_2001}.  DCM allows the
measurement of individual nanomagnets as a function of applied
external field in controlled orientations and provides information on
the saturation magnetization, anisotropy, and the switching behavior.
The technique has been recently used to measure both normal and
superconducting mesoscopic rings
\cite{bleszynski-jayich_persistent_2009, jang_observation_2011},
individual ferromagnetic nanostructures
\cite{banerjee_magnetization_2010, weber_cantilever_2012,
  buchter_reversal_2013}, and the skyrmion phase in a single
nanomagnet \cite{mehlin_stabilized_2015}.  We develop a simple
analytical model for the DCM of magnetic nanostructures, as well as a
numerical framework applicable to a broad range of nanomagnetic
samples.  Using these tools to guide our interpretation of the data,
we find evidence for stable low-field flux-closure configurations and
gain insight on the sequence of the magnetization reversal process.
We note that the applicability of our numerical DCM model is not
limited to ferromagnetic systems and could form the basis for
simulating and interpreting DCM measurements in samples with a variety
of complex magnetic configurations.
%
%
%
%
%
\section{Samples}
\label{sec:samples}
The samples in this study are chosen because of their similarity to
idealized ferromagnetic nanotubes, which have been the subject of
extensive theoretical modelling \cite{hertel_magnetic_2004,
  escrig_phase_2007, escrig_effect_2007, landeros_reversal_2007,
  landeros_equilibrium_2009, chen_magnetization_2010,
  landeros_domain_2010, chen_magnetization_2011, yan_chiral_2012}.
The fabrication process and choice of material facilitate smooth
sample surfaces, a comparatively large saturation magnetization, and
avoid magneto-crystalline anisotropy \cite{ruffer_anisotropic_2014,
  schwarze_magnonic_2013}. These properties yield strong nanomagnets,
whose stable magnetization configurations are determined by their
designed geometry, rather than by defects or geometrical
imperfections.

\begin{figure}[t]
	\centering
	\includegraphics[width=0.90\columnwidth]{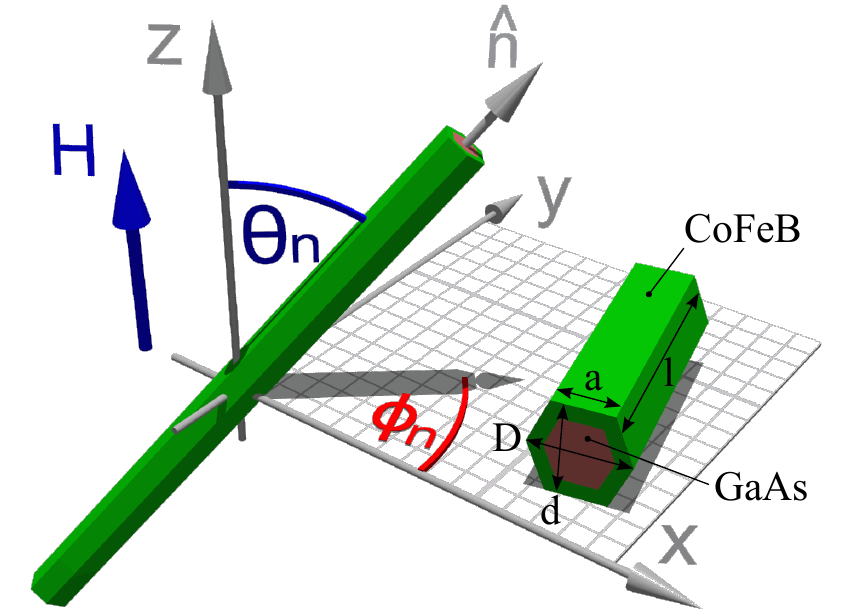}\\[2ex]
	\includegraphics[width=0.90\columnwidth]{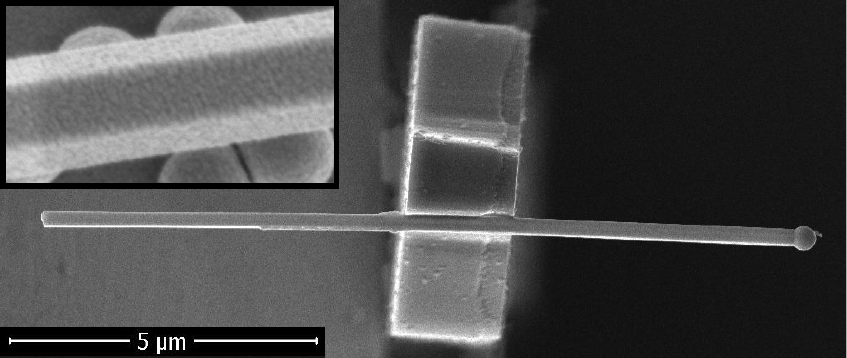}
	\caption{Top: Illustration of the sample structure and
          definition of coordinates. Bottom: SEM image of a CoFeB
          nanotube attached to the end of a Si cantilever (the long
          axis of the cantilever is perpendicular to the image plane). Inset: Close-up of a nanotube.}
	\label{fig:NT_structure}
\end{figure}

The CoFeB nanotubes consist of a non-magnetic GaAs core surrounded by
a magnetic CoFeB shell with a hexagonal cross-section, as sketched in
Fig.~\ref{fig:NT_structure}.
To fabricate the samples, GaAs nanowires are grown on a Si (111)
substrate using Ga droplets as catalysts by molecular beam epitaxy
\cite{ruffer_anisotropic_2014}.  Then CoFeB is magnetron-sputtered on
the nanowires, producing a homogeneously thick ($t=30 \pm
\SI{2}{\nano\meter}$), amorphous shell, avoiding magneto-crystalline
anisotropy in the samples \cite{hindmarch_interface_2008}.  The
saturation magnetization of planar CoFeB films of similar thickness as for the present nanotubes has
been measured to be $\mu_0M_s= 1.80$ T \cite{ulrichs_magnonic_2010}.
The resulting nanotubes are typically over 10 $\si{\micro\meter}$ long
and about 250 $\si{\nano\meter}$ in diameter. The dimensions of the
individual nanotubes used in this study are summarized in
Tab.~\ref{tab:tube_dimensions} as determined using scanning electron
microscopy (SEM), cf.\ Fig.~\ref{fig:NT_structure} for a
representative image. Note, that one end of the nanotubes is
terminated by the Ga droplet from the nanowire growth process, covered
with CoFeB, while at the other end the sample has been broken off the
substrate.  As a result, the end of the tubes are not -- in general --
capped by an open and perfectly flat end.  Nevertheless, SEM images
reveal continuous and defect-free tubes, whose surface roughness is
better than \SI{3}{\nano\meter} \cite{ruffer_anisotropic_2014}.  This
near perfection is in contrast to the Ni nanotubes studied by DCM in
Weber et al.\ \cite{weber_cantilever_2012}, which contained a
peak-to-peak roughness on the order of \SI{10}{\nano\meter}.  Buchter
et al.\ \cite{buchter_reversal_2013} showed that the unintentional
roughness of these Ni nanotubes likely made them different enough from
idealized ferromagnetic nanotubes to result in a magnetization
reversal process unlike that predicted by theory.
\begin{table}
  \caption{Dimensions of the measured nanotubes, quantities are defined in Fig.~\ref{fig:NT_structure}. An error of $\pm \SI{0.02}{\micro\meter}$ is estimated for $D$, $d$ and $a$.}
\label{tab:tube_dimensions}
\begin{ruledtabular}
  \begin{tabular}{cccccc} Config. & $l$ ($\si{\micro\meter}$) & $D$ ($\si{\micro\meter}$) & $d$ ($\si{\micro\meter}$) & $a$ ($\si{\micro\meter}$) & $V$ ($10^{-19}\,\si{\meter}^3$)\\
    1 & $10.3 \pm 1.0$ & $0.26$ & $0.26$ & $0.12$ & $2.3 \pm 0.6$ \\
    2 & $12.6 \pm 0.1$ & $0.27$ & $0.24$ & $0.15$ &  $2.8 \pm 0.5$ \\
    3 & $12.0 \pm 0.1$ & $0.25$ & $0.24$ & $0.11$ &  $2.4 \pm 0.4$ \\
\end{tabular}
\end{ruledtabular}
\end{table}

\section{Experimental Setup}
\label{sec:setup}
\begin{figure}[t]
	\centering
        \includegraphics[width=.90\columnwidth]{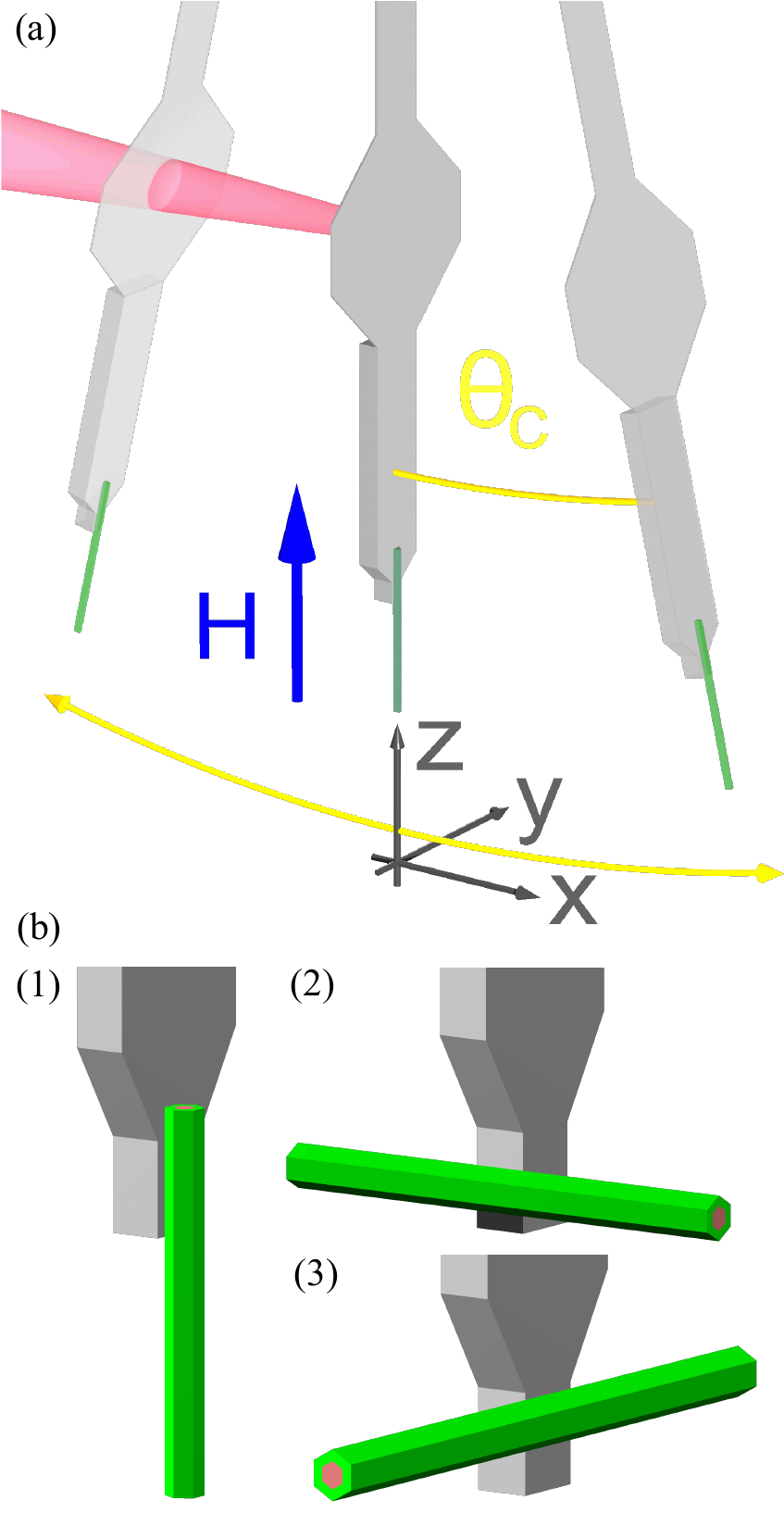}
	\caption{(a) Schematic diagram of the experimental setup and
          (b) the three orientations of the nanotubes with respect to the
          magnetic field, which we label configuration~1, 2, and
          3.}
	\label{fig:DCMsetup}
\end{figure}

DCM involves a measurement in an externally applied magnetic field of
the mechanical resonance frequency of a cantilever, to which the
nanomagnet of interest has been attached.  By using ultra-soft,
single-crystal Si cantilevers, we achieve a sufficiently high
sensitivity to probe the magnetic states of single nanotubes.  In
order to carry out such measurements, individual CoFeB nanotubes are
glued to the end of a cantilever with epoxy (Gatan G1) using
micro-manipulators under a customized optical microscope
\cite{weber_cantilever_2012}.  Three orientations of the nanotubes
relative to the applied magnetic field are prepared, as depicted in
Fig.~\ref{fig:DCMsetup} and labelled configurations 1, 2 and 3.  The
error of the actual nanotube orientations relative to the desired
orthogonal orientations is $\pm \SI{10}{\degree}$.
\begin{table}
  \caption{Properties of the cantilevers used for the experiments.}
\label{tab:catilever_properties}
\begin{ruledtabular}
\begin{tabular}{ccccc}
Config. & $l_e$ ($\si{\micro\meter}$)& $f_0$ ($\si{\hertz}$) & $k_0$ ($\si{\micro\newton / \meter}$) & $Q$\\
1 & $108.2 \pm 0.3 $& 2191.0 & $48.7 \pm 4.5 $& $33 \cdot 10^3$ \\
2 & $107.8 \pm 0.3 $& 2211.9 & $48.5 \pm 4.7 $& $41 \cdot 10^3$ \\
3 & $107.8 \pm 0.3 $& 2107.8 & $40.3 \pm 3.0 $& $36 \cdot 10^3$ \\
\end{tabular}
\end{ruledtabular}
\end{table}

The cantilevers used here are about \SI{150}{\micro\meter} long,
\SI{3.5}{\micro\meter} wide and \SI{0.12}{\micro\meter} thick with a
mass-loaded end being \SI{18}{\micro\meter} long,
\SI{4.9}{\micro\meter} wide and \SI{1.7}{\micro\meter} thick.  Their
fundamental mechanical modes have quality factors $Q$ around $3\cdot
10^{4}$, resonance frequencies $f_0$ between 2 and 3 kHz and spring
constants $k_0$ around \SI{50}{\micro\newton/\meter} under measurement
conditions, i.e.\ in a vacuum chamber with a pressure below $10^{-6}$
mbar in a cryostat with $T = 4.2$ K.  A superconducting
magnet built into the cryostat allows the application of an external
magnetic field up to $\pm 6$ T parallel to the cantilever axis
($\mathbf{\hat{z}}$ axis).  The deflection of the cantilever (along $\mathbf{\hat{x}}$)
is measured by a fiber interferometer \cite{rugar_improved_1989} with
$\SI{100}{\nano\watt}$ of $\SI{1550}{\nano\meter}$ laser light focused
onto a \SI{11}{\micro\meter} wide paddle near the mass-loaded end of
the cantilever.  A piezo-electric actuator mechanically drives the
cantilever at its resonance frequency with a constant oscillation
amplitude of $\SI{40}{\nano\meter}$, using a feed-back loop
implemented by a field-programmable gate array.  This process of
self-oscillation enables the fast and accurate extraction of the
resonance frequency from the cantilever deflection signal.

\section{Dynamic cantilever magnetometry}
\label{sec:DCM}
The energy of the magnet-on-cantilever system can be described by the
sum of a mechanical energy term, related to the cantilever
(approximated here as a simple harmonic oscillator), and a magnetic
energy term, related to the attached sample:
\begin{equation}
  E = \frac{1}{2} k_0 (l_e \theta_c)^2 + E_m , 
\label{eq:Etotal}
\end{equation}
where $k_0$ is the spring constant, $l_e$ is the effective length of
cantilever's fundamental mode, $\theta_c$ is the angle of the
cantilever free-end with respect to $\mathbf{H}$, and $E_m$ is the
magnetic energy.  Given that the Si cantilever and the epoxy used to
attach the sample have no magnetic response, the magnetic energy
depends only on the properties of the attached nanomagnet.  As shown
in Fig.~\ref{fig:DCMsetup}, $\mathbf{H}$ sets $\mathbf{\hat{z}}$ of
our coordinate system, while $\mathbf{\hat{y}}$ is coincident with the
cantilever's axis of rotation.  Therefore the measured cantilever
deflection $\theta_c$ depends on the component of the torque along
$\mathbf{\hat{y}}$, which is given by $\tau_y = -\partial E / \partial
\theta_c$.  Since $\theta_c \ll 1^{\circ}$ during the measurement
(i.e.\ $x/l_e \ll 1$, where $x = l_e \theta_c$ is the position of the
cantilever's free-end), we expand $E_m$ as a function of $\theta_c$
around $\theta_c = 0$.  Keeping only terms up to first order in
$\theta_c$, we find:
\begin{equation}
  \tau_y = - \left(\left . \frac{\partial E_m}{\partial\theta_c} \right |_{\theta_c=0}\right) - \left[k_0 l_e^2 + \left(\left . \frac{\partial^2 E_m}{\partial\theta_c^2} \right |_{\theta_c=0}\right)\right]\theta_c ,
\label{eq:TorqueExpandApp}
\end{equation}
where $\left.\frac{\partial
    E_m}{\partial\theta_c}\right|_{\theta_c=0}$ and
$\left.\frac{\partial^2E_m}{\partial\theta_c^2}\right|_{\theta_c=0}$
are the first and second derivatives of the magnetic energy with
respect to $\theta_c$ at the cantilever's equilibrium angle.  The
equation of motion for this harmonic oscillator is
\begin{equation}
  m_e \ddot{x} + \Gamma \dot{x} = \tau_y/l_e,
\label{eq:Emotion}
\end{equation}
where $m_e$ is the effective mass of the cantilever, and $\Gamma$ is the
cantilever's mechanical dissipation.  We then see that the first
term in (\ref{eq:TorqueExpandApp}) produces a constant deflection of
the cantilever, while the term proportional to $\theta_c$ determines
the cantilever's spring constant:
\begin{equation}
 \begin{split}
  m_e \ddot{\theta_c} + \Gamma\dot{\theta_c} +\left[k_0 + \frac{1}{l_e^2}\left( \left . \frac{\partial^2E_m}{\partial\theta_c^2}\right|_{\theta_c=0}\right)\right]\theta_c =\\
		-\frac{1}{l_e^2}\left( \left . \frac{\partial E_m}{\partial\theta_c}\right|_{\theta_c=0}\right).
 \end{split}
\label{eq:Emotion2}
\end{equation}
Conventional static measurements of cantilever magnetometry keep track
of the constant deflection term, while the DCM measurements presented
here follow the change in the cantilever spring constant.  Using this
equation of motion, we solve for the cantilever's frequency shift
$\Delta f = f - f_0$, where $f$ is the measured resonance frequency
and $f_0$ is the resonance frequency at $H = 0$ (see Mehlin et al.\
for full derivation \cite{mehlin_stabilized_2015}):
\begin{equation}
  \Delta f = \frac{f_0}{2 k_0 l_e^2}\left( \left.\frac{\partial^2E_m}{\partial\theta_c^2}\right|_{\theta_c=0}\right).
  \label{eq:deltaf}
\end{equation}
Note that we have neglected a term in $\Delta f$ which depends on the
cantilever dissipation $\Gamma$, which for the cantilevers and samples
used here is negligible compared to the magnetic anisotropy term.
Measurements of $\Delta f$ thus reveal the curvature of the magnetic
energy with respect to sample rotations about the cantilever
oscillation axis.  By mounting the sample in various configurations,
we investigate the energy curvature about the various rotation axes.

\section{Analytical Model}
\label{sec:magnetization}

\subsection{An Idealized Single-domain Magnet}
\label{subsec:SW}

In order to establish a framework from which to interpret our DCM
measurements, we begin by modelling our ferromagnetic nanotubes as
idealized single-domain magnets.  In this simplified model, the
magnetization is uniform throughout the magnet and rotates in unison.
Its direction is determined by its magnetostatic energy, which we
reduce to the Zeeman energy and the anisotropy energy.  Given the
polycrystallinity and large aspect ratio of the nanotubes (roughly
30:1), the latter is dominated by shape anisotropy.  In literature
\cite{aharoni_introduction_2000, skomski_permanent_1999}, magnetic
anisotropy of small magnetic particles has frequently been addressed
by working within the Stoner-Wohlfarth (SW) approximation
\cite{stoner_xcvii._1945}.  For simplicity, particles are modelled as
uniformly magnetized, prolate ellipsoids of revolution, with a
demagnetizing field $\mathbf{H}_{\text{dm}}= -D\circ \mathbf{M}$ produced by
the magnetization $\mathbf{M}$, where $\mathbf{H}_{\text{dm}}$ turns out to
be homogeneous.  $D$ is a tensor consisting of the diagonal elements
$D_x$, $D_y$ and $D_z$, the demagnetization factors, which describe
the anisotropy due to the shape of the particle.  It can be shown
that, in a uniform applied field, the magnetization of a single-domain
particle of arbitrary shape behaves precisely as that of a suitably
chosen ellipsoid \cite{brown_effect_1957}.  By applying this
generalization, one can derive the magnetometric demagnetization
factors and therefore the magnetic behavior for any arbitrary
single-domain magnet.  Note, however, that for non-ellipsoids the
homogeneity of $\mathbf{H}_{\text{dm}}$ within the magnet is not preserved.

In order to calculate demagnetization factors for our ferromagnetic
nanotubes, we first approximate the nanotubes as hollow cylinders
\cite{beleggia_demagnetization_2009}, ignoring the hexagonality of
their cross-section.  This may be justified by the large aspect ratio
between length and diameter of the tubes and leads to demagnetization
factors of $D_x = D_y \approx 0.498$ and $D_z \approx 0.004$ for the
tube axis parallel to the $\mathbf{\hat{z}}$ axis.  This implies that we consider
only uniaxial anisotropy, which can also be described by a unit vector
$\mathbf{\hat{n}}$ along the tube axis and an effective demagnetization factor
$D_u = D_z - D_x$ \cite{aharoni_introduction_2000,skomski_permanent_1999}.  Effects of the hexagonal cross-section, leading to
deviations from the uniaxial description, are then discussed in section
\ref{sec:results}.

The SW model describes the ferromagnetic nanotubes accurately for
large applied external fields, in which the nanotubes are forced into
uniform magnetization configurations.  However, using this model for
low applied fields, where non-uniform magnetic configurations exist,
is a strong simplification with limited validity.  For this regime,
discussed in section \ref{sec:numerical}, we employ micromagnetic
simulations to describe the behavior of the magnetization
configurations and the resulting DCM signal.  Nevertheless, comparing
measurements to the SW model serves as an indicator of the extent to
which magnetic configurations are uniform and magnetization reversal
is coherent.

\subsection{SW Approximation Applied to DCM}
\label{subsec:SWappliedToDCM}

In the SW model, $E_m$ and the magnetic history of the magnet
determine the equilibrium magnetization $\mathbf{M}$.  $E_m$ in turn
depends on the external field $\mathbf{H}$ and the magnet's
properties.  In order to calculate $E_m$, we define the different
orientations of the nanotube in our coordinate system, as depicted in
Fig.~\ref{fig:NT_structure}, with the spherical coordinates $\theta_n$
and $\phi_n$.  Likewise, we describe the net magnetization
$\mathbf{M}$ with the angles $\phi_m$ and $\theta_m$, so that,
\begin{equation}
  \mathbf{M} = M_s \begin{pmatrix} \sin\theta_m\cos\phi_m \\ \sin\theta_m\sin\phi_m \\ \cos\theta_m \end{pmatrix},
\label{eq:magnetization}
\end{equation}
where $M_s$ is a constant.  In this approximation, the magnetic energy
consists of a Zeeman term and an anisotropy term:
\begin{equation}
  E_m =  -\mu_0 V \mathbf{M}\cdot\mathbf{H} + \frac{1}{2}\mu_0 
  V D_u (\mathbf{M} \cdot \mathbf{\hat{n}})^2,
\label{eq:magneticEnergy}
\end{equation}
where $\mu_0$ is the vacuum permeability and $V$ is the volume of the
nanotube.  Therefore, $E_m$ depends on $\mathbf{H}$, $\phi_m$,
$\theta_m$, $\phi_n$, and $\theta_n$.  In order to determine the
behavior of $\mathbf{M}$ in our experiment, we introduce the
oscillating cantilever.  Oscillation of the cantilever amounts to a
rotation of the nanotube orientation $\mathbf{\hat{n}}$ about $\mathbf{\hat{y}}$.  This
process introduces an additional $\theta_c$ dependence to the nanotube
orientation and thus also to $E_m$.

\begin{figure*}[tb]
	\centering
	\includegraphics[width=1.8\columnwidth]{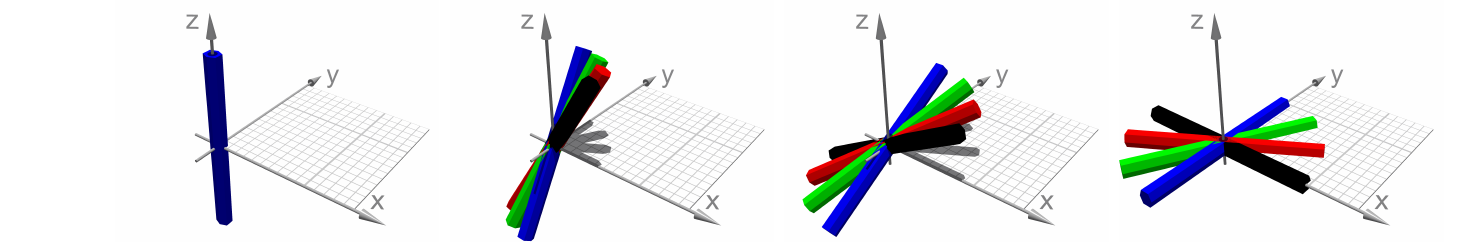}\\
	\includegraphics[width=1.8\columnwidth]{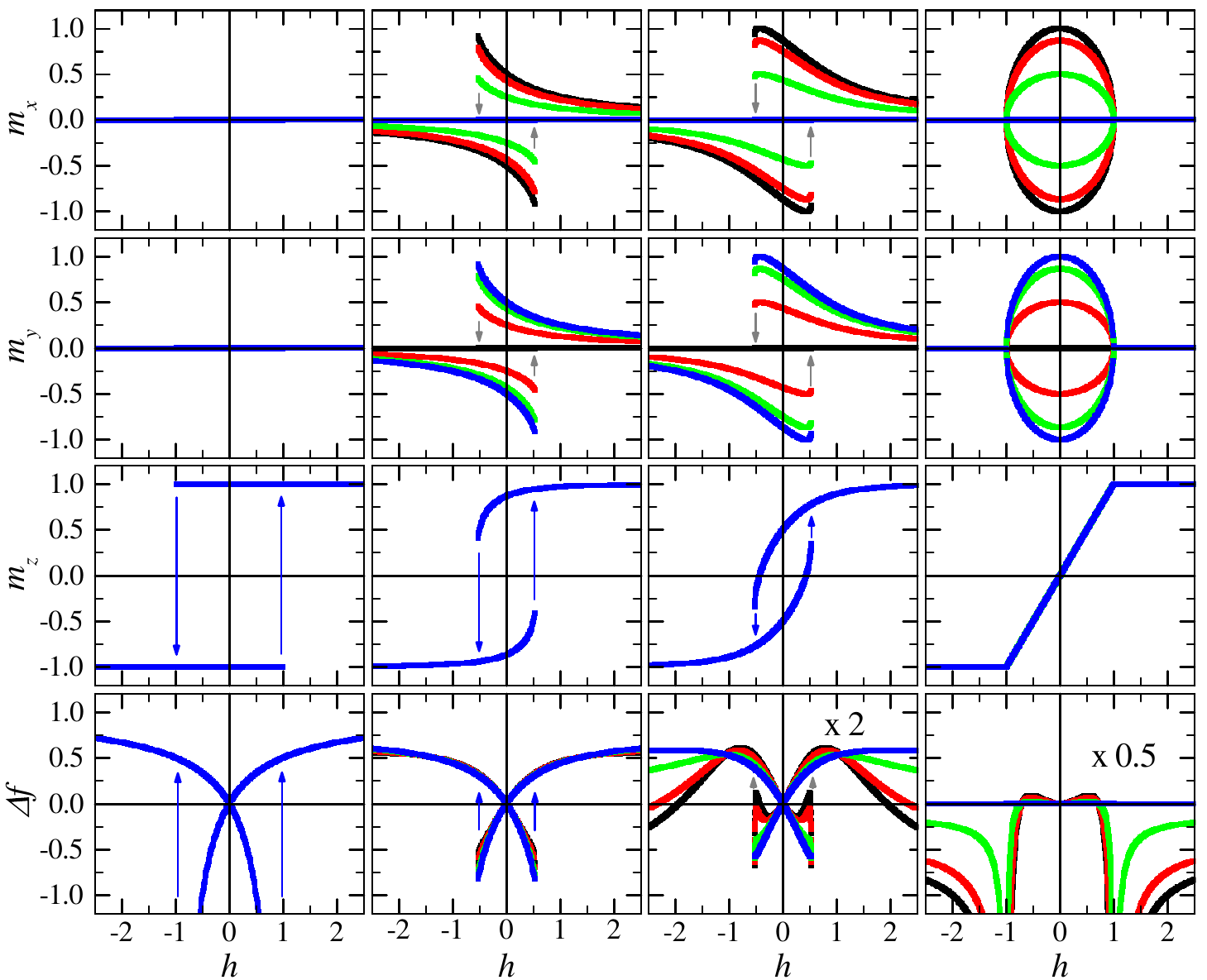}
	\caption{Components of $\mathbf{M}$ normalized by $M_s$ (first
          3 rows) and $\Delta f$ normalized by $\frac{f_0 \mu_0 V
            M_s^2}{2 k_0 l_e^2}$ (last row) vs.\ normalized magnetic
          field $h=-\frac{H}{M_s D_u}$ for different orientations of
          the nanotubes. $\theta_n$ is increased from
          $\SI{0}{\degree}$ in the first column by $\SI{30}{\degree}$
          per column up to $\SI{90}{\degree}$. $\phi_m$ is changed in
          the same steps, given by the different, color-coded graphs
          within each column. Arrows indicate switching of the
          magnetization.}
	\label{fig:greatgraph}
\end{figure*}

Since the microscopic processes in ferromagnetic nanotubes are
expected to be much faster than the cantilever resonance frequency
\cite{landeros_domain_2010, schwarze_magnonic_2013, yan_chiral_2012},
the magnetization of the nanotube can always be assumed to be in its
equilibrium orientation.  We can therefore solve for $\phi_m$ and
$\theta_m$, by fulfilling the following minimization conditions for
$E_m$:
\begin{eqnarray}
  \frac{\partial E_m}{\partial \phi_m} = \frac{\partial E_m}{\partial
    \theta_m} & = & 0; \\
  \frac{\partial^2 E_m}{\partial \phi_m^2}, \; \frac{\partial^2 E_m}{\partial \theta_m^2} & > & 0.
\label{eq:magangles}
\end{eqnarray}
Since $\theta_c \ll 0.1^{\circ}$, we approximate the solutions for
$\phi_m$ and $\theta_m$ by considering only terms up to first order in
$\theta_c$:
\begin{eqnarray}
\label{eq:magangles1}
\phi_m(\theta_c) & \approx \phi_m(0) + \left.\frac{\partial\phi_m}{\partial\theta_c}\right|_{\theta_c = 0}\cdot \theta_c \\
\theta_m(\theta_c) & \approx \theta_m(0) + \left.\frac{\partial\theta_m}{\partial\theta_c}\right|_{\theta_c = 0}\cdot \theta_c.
\label{eq:magangles2}
\end{eqnarray}
As expected, once solutions of this form are found, we see that the
equilibrium angles of the magnetization for $\theta_c = 0$ give the
solutions already known from the SW model
\cite{morrish_physical_2001}.  $\phi_m(0) = \phi_n$, such that the
azimuthal orientation of the magnetization always follows the
azimuthal orientation of the magnet.  In other words, the
magnetization is constrained to the plane defined by the magnet's
uniaxial anisotropy axis and the direction of the magnetic
field. $\theta_m(0)$ is given by the arctangent of a solution to a
quartic equation.  Either one or two of the four possible solutions
for $\theta_m(0)$ are real and minimize $E_m$.  When there are two
physical solutions, the system allows for magnetic hysteresis.  In the
first three rows of Fig.~\ref{fig:greatgraph}, we plot the components
of the resulting equilibrium magnetization $\mathbf{M}$ normalized to
$M_s$ as function of the reduced magnetic field $h=-\frac{H}{M_s D_u}$
for $\theta_c = 0$ and several orientations $\phi_n$ and $\theta_n$.

Solutions for $\phi_m(\theta_c)$ and $\theta_m(\theta_c)$ in the form
of (\ref{eq:magangles1}) and (\ref{eq:magangles2}) are then replaced
into the expression for $E_m$.  By taking the second derivative of
$E_m$ with respect to $\theta_c$ and applying (\ref{eq:deltaf}), we
find the corresponding frequency shift measured by DCM.  We implement
the above procedure in a \textit{Mathematica} script, which allows us
to determine the magnetization and corresponding frequency shift for
any orientation of the nanotube ($\phi_n$, $\theta_n$) in any applied
magnetic field.  This approach is similar to that described by others
\cite{harris_high_2000,
  straver_cantilever_2004,kamra_theoretical_2014}.  The last row of
Fig.~\ref{fig:greatgraph} shows the corresponding frequency shift
measured by DCM, $\Delta f$, as function of $h$ and normalized to
$\frac{f_0 \mu_0 V M_s^2}{2 k_0 l_e^2}$.

In general, a positive value of $\Delta f$ implies that
$E_m(\theta_c)$ is in a local minimum with respect to $\theta_c$, and
therefore changing the cantilever angle increases $E_m$.  In other
words, given the alignment of the nanotube's net magnetization, its
physical orientation in the $xz$-plane is energetically
favorable. Therefore, the cantilever experiences a spring-like
magnetic restoring force, which stabilizes this orientation. This
``magnetic spring'' produces an effective hardening of the cantilever spring
constant, reflected as an increase in its resonant frequency.  On the
other hand, a negative value of $\Delta f$ corresponds to a local
maximum in $E_m(\theta_c)$.  That is, given the alignment of the
nanotube's net magnetization, its physical orientation in the
$xz$-plane is energetically unfavorable.  The cantilever thus
experiences an anti-restoring force, resulting in an effective softening of the cantilever spring constant and thus a reduction
in its resonance frequency.
\subsection{The High-field Limit}
\label{subsec:magnetizationHigh}
Regardless of the nanotube's orientation, for large fields, i.e.\ $H
\gg \left|D_u M_s\right|$, $\mathbf{M}$ is forced to be parallel to
$\mathbf{H}$. By applying this limit to (\ref{eq:magneticEnergy}) and
solving (\ref{eq:deltaf}), we find that $\Delta f$ approaches a
horizontal asymptote given by:
\begin{equation}
  \Delta f = \frac{f_0 \mu_0 V}{2 k_0 l_e^2} M_s^2 D_u \left( \sin ^2\theta_n \cos^2\phi_n - \cos^2\theta_n \right ).
\label{eq:limit}
\end{equation}
In this limit, $\Delta f$ is a measure of the anisotropy energy of the
nanomagnetic particle, multiplied by a factor depending on its
orientation relative to $\mathbf{H}$.  In short, by forcing
$\mathbf{M}$ and $\mathbf{H}$ to be parallel, the cantilever
oscillation only probes the curvature of the second term in the
magnetic energy shown in (\ref{eq:magneticEnergy}).  For example, as
can be seen from (\ref{eq:limit}) and the last row of
Fig.~\ref{fig:greatgraph}, at high field, a nanotube oriented along
$\theta_n=0$ will approach an equal and opposite $\Delta f$ as when it
is oriented along $\phi_n=0$ and $\theta_n=\SI{90}{\degree}$.  When
the anisotropy axis $\mathbf{\hat{n}}$ of an idealized nanotube is coincident
with the axis of cantilever oscillation $\mathbf{\hat{y}}$ ($\phi_n = \theta_n
= \SI{90}{\degree}$), $\Delta f = 0$.  The cantilever frequency is
unaffected in this geometry, because the idealized cylindrical
nanotube is symmetric about $\mathbf{\hat{n}}$.  Since $\mathbf{\hat{n}} \parallel \mathbf{\hat{y}}$, the magnetic energy has no curvature along $\theta_c$.

\subsection{Intermediate Fields}
\label{subsec:intermediateFields}
At intermediate fields, we can understand the relationship between
$\Delta f (H)$ curves and the SW magnetization curves by considering
the limiting orientations. The first column of
Fig.~\ref{fig:greatgraph} shows the case of a nanotube with its easy
axis aligned along $\mathbf{H}$ ($\theta_n = 0$).  The magnetization
$M_z$ executes the expected square hysteresis loop as a function of
$H$, while $M_x=M_y=0$.  As $H$ is swept down from high fields,
$\mathbf{M}$ is parallel to $\mathbf{H}$, making the physical
orientation of the nanotube in the $xz$-plane energetically
optimal. As a result, $\Delta f > 0$.  Upon crossing $H = 0$, however,
the direction of $\mathbf{H}$ inverts and $\mathbf{M}$ becomes
anti-parallel to $\mathbf{H}$, making the nanotube's orientation
energetically unfavorable.  As a result, $\Delta f < 0$.  Once the
coercive field is reached and $\mathbf{M}$ again switches into a
parallel orientation with respect to $\mathbf{H}$, the frequency shift
also switches sign giving $\Delta f > 0$.

The last column of Fig.~\ref{fig:greatgraph} shows the case of a
nanotube with its easy axis aligned perpendicular to $\mathbf{H}$
($\theta_n = \SI{90}{\degree}$).  As $H$ is increased across zero, the
magnetization $\mathbf{M}$ rotates coherently from alignment with
$-\mathbf{\hat{z}}$ through the $xy$-plane to $+\mathbf{\hat{z}}$ without hysteresis. As
$H$ is swept down from high fields, $\mathbf{M}$ is parallel to
$\mathbf{H}$ and perpendicular to the anisotropy axis, making the
physical orientation of the nanotube energetically unfavorable and
giving $\Delta f < 0$.  Once $H$ is reduced enough that $\mathbf{M}$
has tilted closer to the $xy$-plane than to $\mathbf{\hat{z}}$ (i.e.\ $\theta_m
> \SI{45}{\degree}$), the orientation of the nanotube becomes
energetically favorable and $\Delta f > 0$.  As $H$ decreases across
zero, the behavior is symmetric.

Intermediate orientations depicted in the middle columns of
Fig.~\ref{fig:greatgraph} show the effects of arbitrary alignments of
the nanotube with respect to $\mathbf{H}$.  In general, alignments
between $\mathbf{\hat{n}} \parallel \mathbf{H}$ and $\mathbf{\hat{n}} \perp \mathbf{H}$
reduce the expected $|\Delta f|$.  Even a slight misalignment from the
$\mathbf{\hat{n}} \perp \mathbf{H}$ case, introduces hysteresis to the
magnetization loop, since the energetic symmetry of $\pm\mathbf{\hat{n}}$ is
broken by $\mathbf{H}$.

Note that for all orientations at $H = 0$, $\Delta f = 0$.  Indeed,
the effects that we observe all arise due to the interaction between
the magnetization of our nanotube and the externally applied field
$\mathbf{H}$.  A curvature in $E_m$ exists with respect to $\theta_c$
only because $\theta_c$ changes the nanotube's orientation relative to
$\mathbf{H}$; when $H=0$, these interactions vanish.  As a result, the
sensitivity of DCM becomes progressively worse as $\mathbf{H}$
approaches zero, at which point the technique is completely
insensitive.  Jang et al.\ describe a variation on DCM, known as
phase-locked cantilever magnetometry (PLCM), with the additional
application of an AC magnetic field in order to overcome this
limitation \cite{jang_phase-locked_2011}.

\subsection{The Low-field Limit}
\label{subsec:lowFields}
For low applied magnetic fields, i.e.\ $H \ll \left|D_u M_s\right|$, shape
anisotropy dominates the magnetic energy in the SW model and ensures
that the net magnetization $\mathbf{M}$ remains either parallel or
antiparallel to the nanotube axis $\mathbf{\hat{n}}$.  In this case, the
cantilever oscillation only probes the curvature of the first term in
the magnetic energy shown in (\ref{eq:magneticEnergy}).  Applying this
limit to (\ref{eq:magneticEnergy}) and solving (\ref{eq:deltaf}), we
find:
\begin{equation}
  \Delta f = \frac{f_0 \mu_0 V}{2 k_0 l_e^2} H M_s
  \cos{\theta_n} = \frac{f_0 \mu_0 V}{2 k_0 l_e^2} H M_z.  
\label{eq:lowFieldLimit}
\end{equation}
This low-field regime constitutes a special case, since this
expression allows the direct determination of $M_z$ from measurements
of $\Delta f$.  By solving (\ref{eq:lowFieldLimit}) for $M_z$, at low
field we have:
\begin{equation}
  M_z = \frac{2 k_0 l_e^2}{f_0 \mu_0 V H} \Delta f.
\label{eq:MagVsDeltaf}
\end{equation}
Despite the fact that non-uniform magnetization configurations are
likely present in the nanotubes at low field, these equations allow us
to analyze DCM data and extract an effective magnetization that
describes the behavior of an equivalent SW magnet.  Such analysis
allows us to see the extent to which the magnetic configuration within
the nanotube is uniform and rotates coherently.
\section{Numerical Calculations}
\label{sec:numerical}
The limitations of any model for DCM based on the SW approximation are
clear: multi-domain magnets or magnets having non-uniform
magnetization configurations cannot be described.  For this reason, we
carry out hybrid finite/boundary element simulations using the
software package \textit{Nmag} \cite{fischbacher_systematic_2007},
allowing us to model the magnetization distribution within a
nanotube. From these simulations we then calculate the DCM frequency
shift.
\subsection{Simulation of Magnetization Configurations}
\label{subsec:magSimulations}
\textit{Nmag} determines the magnetization distribution step-by-step
for each external field value by numerically solving the
Landau-Lifschitz-Gilbert equation.  We model our nanotubes as
perfectly hexagonal tubes with an inner diameter of
\SI{190}{\nano\meter} and a thickness of \SI{30}{\nano\meter}, with an
alignment to the external magnetic field that can be freely
chosen. Our computational capacity limits us to tubes of
\SI{1.5}{\micro\meter} length, when keeping the mesh cell size below
about \SI{8}{\nano\meter}.

By employing periodic boundary conditions (PBCs), this length
limitation can be overcome.  This choice, however, implies that the
effects of the tube ends are not included in the simulations.  PBCs
preclude the modelling of end-states such as vortex configurations,
which nucleate at the tube ends and which have been predicted to
initiate magnetization reversal
\cite{hertel_magnetic_2004,landeros_equilibrium_2009}.

We therefore carry out two types of simulations: the first with
\SI{1.5}{\micro\meter} long tubes without PBCs; and the second with
200 repetitions of a \SI{150}{\nano\meter}-long segment with PBCs.  We
set the exchange coupling constant to $A=\SI{28}{\pico\ampere/\meter}$
\cite{bilzer_study_2006} and use the saturation magnetization $M_s$ as
the only fit parameter.  The initial value for $M_s$ is extracted from
fits to our high-field DCM measurements using our analytical SW model.

\subsection{Simulation of DCM Frequency Shift}
\label{subsec:DCMSimulations}
We employ the following procedure in order to calculate the DCM
frequency shift from the \textit{Nmag} simulations:
\begin{enumerate}
\item For each value of the external magnetic field $H$, we calculate
  the magnetization configuration with \textit{Nmag} for the
  cantilever in its equilibrium orientation $\theta_c = 0$.
\item We then calculate the magnetization $\mathbf{m}_i$ at the
  centroid $\mathbf{r}_i$ of each tetrahedral mesh element $i$.  We
  use a Shepard weighting $w_j(\mathbf{r}_j) =
  \left|\mathbf{r}_i-\mathbf{r}_j\right|^{-2}$, where the indices
  refer to the $j$th vertex of the $i$th tetrahedron, to
  obtain $$\mathbf{m}_i =
  \frac{\sum_{j}{w_j\mathbf{m}_j}}{\sum_{j}{w_j}}.$$ This calculation
  is necessary, since \textit{Nmag} determines the magnetization at
  the vertices of the tetrahedra only.
\item We calculate the magnetic moment $\mathbf{\mu}_i$ of each
  tetrahedron via $\boldsymbol{\mu}_i = \mathbf{m}_iV_i,$ with the
  volume $V_i$ of the corresponding tetrahedron.  The total magnetic
  moment of the tube is then $\boldsymbol{\mu}_{tube} =
  \sum_{i}{\boldsymbol{\mu}_i}$.  The resultant magnetic torque on the
  tube along the cantilever's axis of rotation is given
  by $$\tau_{m,y} = (\boldsymbol{\mu}_{tube} \times
  \mu_0\mathbf{H) \cdot \mathbf{\hat{y}}}.$$
\item Having calculated the magnetic torque on the tube for each value
  of $H$ in the field sweep at $\theta_c = 0$, we now tilt the tube by
  a small but finite angle $\delta \theta_c \leq 0.3^{\circ}$ and
  repeat the three preceding steps.  Now in addition to
  $\tau_{m,y}(0)$, we obtain $\tau_{m,y}(\delta \theta_c)$.
\item We then find the DCM frequency shift using $\tau_{m,y} =
  - \partial E_m / \partial \theta_c$ and (\ref{eq:deltaf}):
\begin{equation}
\begin{split}
  \Delta f &= \frac{f_0}{2k_0l_e^2} \left(-\left.\frac{\partial\tau_{m,y}}{\partial\theta_c}\right|_{\theta_c=0}\right) \\
  &\approx -\frac{f_0}{2k_0l_e^2}\frac{\tau_{m,y}(\delta
    \theta_c)-\tau_{m,y}(0)}{\delta \theta_c}.
\end{split}
\label{eq:Dfmicromag}
\end{equation}
\end{enumerate}
Since the modeled nanotubes are shorter than those measured, for
comparison with the measurements, we scale the calculated $\Delta f$
proportionally with the ratio of volume of the measured and simulated
tubes.  The measured volume, and therefore this ratio, is determined
by measuring the geometry of the nanotubes as discussed in
section~\ref{sec:results}.  After a few iterations altering $M_s$ to
optimize agreement with the experimental DCM data at high field, this
procedure allows us to extract a value for $M_s$.
\section{Experimental Results}
\label{sec:results}
Each CoFeB nanotube is a complex magnetic system consisting of roughly
$10^{10}$ spins.  A variety of spin configurations other than the
simple macro-spin configurations described by the SW approximation can
be expected to occur during a field sweep
\cite{castano_metastable_2003, wang_spin_2005,
  landeros_equilibrium_2009, chen_magnetization_2011,
  ruffer_magnetic_2012, buchter_reversal_2013,
  streubel_magnetic_2014}.  Nevertheless, for high fields, at which
the Zeeman term dominates over interaction terms in the magnetic
energy, treating the system as a single macro-spin is valid.  In the
following, we first analyze this high-field regime and then turn to
low fields, where the SW-model begins to break down.  For that regime,
we rely on micromagnetic simulations for further insight.
\subsection{The High-field Limit}
\label{subsec:highfield}
\begin{figure}[t]
	\centering
	\includegraphics[width=.9\columnwidth]{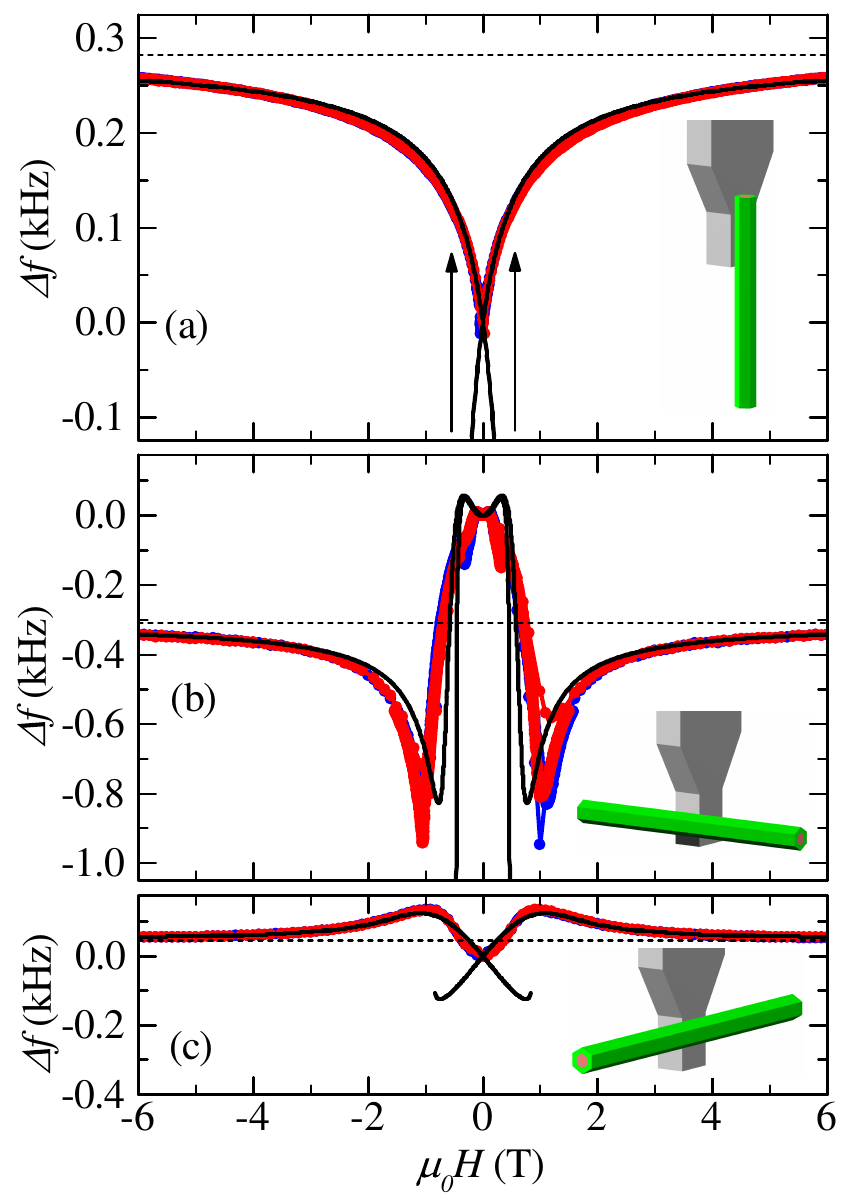}
	\caption{(a), (b), and (c) frequency shift $\Delta f$ vs $\mu_0 H$ for configuration~1, 2, and 3 of the nanotubes, respectively. Red (blue) curves are for field sweeps in positive (negative) direction and black curves are SW-fits using the parameters given in Tab.~\ref{tab:fitparameters}. Dashed lines are horizontal asymptotes.}
	\label{fig:highfield}
\end{figure}

In Fig.~\ref{fig:highfield}, we plot the measured DCM frequency shift
in a field range of $\pm \SI{6}{\tesla}$ for the three different
configurations of the nanotubes as depicted in
Fig.~\ref{fig:DCMsetup}.
Data points taken while $H$ is swept in the positive direction are
red, while points taken for negative sweeps are blue.  To fit the data
with the model developed in section \ref{sec:magnetization} for the
frequency shift of a SW magnet, we use $k_0$ as determined from
thermal noise spectra, $l_e$ from finite element simulations
\cite{montinaro_coupling_2014}, and $V$ approximated from SEM
images. The demagnetization factor $D_u = D_z - D_x$ is calculated by
using the geometry of each nanotube measured from SEM images, approximating
it as a hollow cylinder, and following the method of Beleggia et
al.~\cite{beleggia_demagnetization_2009}.  $f_0$ can be extracted from
the measurements of the cantilever resonance frequency at $H = 0$. The
saturation magnetization $M_s$ is used as a fit parameter.  The
orientation of the nanotubes in the three different configurations,
intended to be
$(\theta_n,\,\phi_n)=(\SI{0}{\degree},\,\SI{0}{\degree})$,
$(\SI{90}{\degree},\,\SI{0}{\degree})$ and
$(\SI{90}{\degree},\,\SI{90}{\degree})$, cannot be expected to be
perfect due to the manual attachment of the tubes to the
cantilever. Therefore, $\theta_n$ and $\phi_n$ are also used as fit
parameters within a $\pm \SI{10}{\degree}$ range of the intended
angles, as estimated from optical and SEM images.  The resulting
curves (black) are plotted with the experimental data in
Fig.~\ref{fig:highfield}.  The relevant cantilever properties in
Tab.~\ref{tab:catilever_properties} and the corresponding fit
parameters are summarized in Tab.~\ref{tab:fitparameters}.  Horizontal
asymptotes are depicted as dashed black lines.
\begin{table}
  \caption{Parameters for fitting $\Delta f$ for the 3 configurations of the CoFeB nanotubes. $D_x$, $D_y$ and $D_z$ are used as fixed parameters, while $M_s$, $\theta_n$ and $\phi_n$ are free parameters.}
\label{tab:fitparameters}
\begin{ruledtabular}
\begin{tabular}{cccccc}
  Config. & $D_x = D_y$ & $D_z$ & $\theta_n$ ($\si{\degree}$) & $\phi_n$ ($\si{\degree}$) & $\mu_0 M_s$ ($\si{\tesla}$) \\
  1 & 0.4977 & 0.0045 & 3 & 90 & $1.29 \pm 0.18$\\
  2 & 0.4981 & 0.0037 & 86 & 5 & $1.21 \pm 0.12$\\
  3 & 0.4981 & 0.0038 & 80 & 90 & $2.5$\\
\end{tabular}
\end{ruledtabular}
\end{table}

Configuration~1, cf.\ Fig.~\ref{fig:highfield} (a), shows a horizontal
asymptotic behavior for large fields, where the asymptote is
approached from lower values.  Data from positive and negative sweep
directions coincide very well and the SW fit gives a good match with
the measurement.  In this orientation, the nanotube behaves like an SW
magnet everywhere except near the low-field magnetization reversal, as
will be discussed in section~\ref{subsec:lowfield}.  From the fit to
the analytical model, we extract a saturation magnetization $\mu_0 M_s
= 1.29 \pm \SI{0.18}{\tesla}$ and orientation angles $\theta_n=
\SI{3}{\degree}$ and $\phi_n= \SI{90}{\degree}$, which lie within the
reasonable range.

Finite element simulations are also carried out according to
section~\ref{subsec:magSimulations} for an optimal high-field fit to
the measurement.  Using the same orientation angles as for the SW
model, the simulations yield $\mu_0 M_s = 1.32 \pm \SI{0.18}{\tesla}$.
Note that the numerical simulations are completely independent of the
calculated demagnetization factors and depend only on the exchange
coupling constant $A$ and the geometry of the nantube.  The high-field
agreement between the SW model and the simulations provides further
confirmation of the extracted value for $M_s$ and the single-domain
behavior of the nanotubes at high fields.

For configuration~2, we find a similarly good agreement between data
and analytical fit function in the high-field regime.  The saturation
magnetization $\mu_0 M_s = 1.21 \pm \SI{0.12}{\tesla}$ extracted from
this measurement is in agreement with that extracted from
configuration~1.  Furthermore, $\theta_n= \SI{86}{\degree}$ and
$\phi_n= \SI{5}{\degree}$ again lie within the reasonable range.  The
corresponding micromagnetic simulations yield $\mu_0 M_s = 1.24 \pm
\SI{0.12}{\tesla}$ for an optimal high-field fit, which is once again
equal to the value extracted from the SW model within our error.  In
this orientation, the behavior of the nanotube magnetization begins to
deviate from the SW model as the SW magnetization begins to coherently
rotate toward the $xy$-plane.  While the qualitative features of
$\Delta f$ displayed by the nanotube and the SW model are the same, as
will be discussed in section~\ref{subsec:lowfield}, the differences
indicate a magnetization not executing an idealized coherent rotation.

The third configuration gives a less conclusive picture.  Although, as
shown in Fig.~\ref{fig:highfield} (c), a fit can reproduce the
behavior of the data for high fields with reasonable values of
$\theta_n$ and $\phi_n$, it results in an anomalously high saturation
magnetization $\mu_0 M_s = \SI{2.5}{\tesla}$.  This value is larger
both than the values extracted in the other orientations and the value
known for a planar thin film CoFeB of $\SI{1.8}{\tesla}$
\cite{ulrichs_magnonic_2010}. The largest source of error in our
determination of $M_s$ comes from our measurement of the nanotube
volume $V$.  $V$ is determined by measuring the outer geometry of the
nanotubes using SEM images and measuring the mean CoFeB shell
thickness from TEMs of representative nanotubes.  The uncertainty in
$V$ results in an error in $M_s$ of nearly $\pm 10\%$.  Nevertheless,
such a measurement uncertainty is not large enough to explain the
anomalously large $M_s$ extracted from configuration~3.

A likely explanation for the failure of this fit is that our
simplified cylindrical model does not take into account magnetic
anisotropies in the plane of the nanotube's hexagonal
cross-section. In fact, for a perfectly aligned tube in this
orientation, the uniaxial anisotropy axis $\mathbf{\hat{n}}$ coincides
with the cantilever oscillation axis $\mathbf{\hat{y}}$.  As a result,
according to our uniaxial model, there should be no variation of $E_m$
with respect to $\theta_c$ and therefore no frequency shift, as shown
by the blue curve in the bottom right graph of
Fig.~\ref{fig:greatgraph}.  Any detected $\Delta f$ stems from either
a misalignment of the nanotube or from deviations of the real sample
from a cylindrically symmetrical SW tube.  In particular, the
hexagonal symmetry of the real sample plays an important role in this
configuration, given that it produces a curvature of $E_m$ about
$\mathbf{\hat{n}}$.  The magnetic energy of the nanotube cannot, in
fact, be fully described by a uniaxial anisotropy, but requires
further axes.  In configurations 1 and 2 the uniaxial model is
appropriate because the anisotropy related to $\mathbf{\hat{n}}$
overwhelms all others in the plane measured by DCM.  In configuration
3, however, anisotropies in the cross-section of the nanotube dominate
and the anisotropy related to $\mathbf{\hat{n}}$ only contributes to
$\Delta f$ in the case of misalignment of the sample.  Since our model
does not take into account the anisotropy in the nanotube
cross-section, it underestimates the effective anisotropy.  As can be
seen from the behavior of the high-field asymptote in
(\ref{eq:limit}), an underestimate of $D_u$ then results in the
extraction of an anomalously large $M_s$ from the fits, just as
observed.  Given the difficulty of disentangling the effects due to
misalignment and additional cross-sectional anisotropies, we abandon a
detailed analysis of configuration 3 using the SW model.

Finite element simulations of $\Delta f (H)$ are carried out in this
configuration and optimized to match measurements at high field.
These numerical calculations yield $\mu_0 M_s = 1.25$ T for a large
but possible misaligment of $\theta_n = 70^{\circ}$ and $\phi_n =
90^{\circ}$.  Nevertheless, modelling this configuration remains
problematic given the measurement's sensitivity to anisotropies
determined by misalignment and the precise cross-section of the
nanotube.  Imperfections in the form of asymmetires in the hexagonal
cross-section along the nanotube's length could cause significant
deviations between the behavior of the real sample and the idealized
model.  In configurations 1 and 2, the effect of such imperfections
cause smaller discrepancies between measurement and model given that
the measured magnetic confinement ($\partial^2 E_m / \partial
\theta_c^2$) is dominated by the large aspect ratio of the nanotubes
rather than their precise shape.  For these reasons, we focus the rest
of our analysis on the experiments carried out in configurations 1 and
2.

Finally, the reason for the discrepancy between the saturation
magnetization determined from configurations 1 and 2, $\mu_0 M_s = 1.3
\pm \SI{0.1}{\tesla}$, and value reported for two-dimensional CoFeB
films of similar thickness, $\SI{1.80}{\tesla}$
\cite{ulrichs_magnonic_2010} remains unclear.  Errors in the
determination of $V$ and $k_0$ are not large enough to explain this
mismatch.  Material degradation through the formation of a outer
oxidation layer is also insufficient to explain the reduction in
saturation magnetization.  About $\SI{15}{\nano\meter}$ of the
$\SI{30}{\nano\meter}$ magnetic shell would have to oxidize in order
account for the difference, while X-ray absorption spectroscopy (XAS)
measurements indicate the presence of an oxide no thicker than 5 nm.
Annular dark field (ADF) scanning transmission electron microscopy
(STEM) of nanotubes produced under identical conditions shows local
variations in the density of the material
\cite{ruffer_anisotropic_2014}, possibly caused by directional
deposition.  We are left to conclude that the reduced saturation
magnetization with respect to planar films is the result of such
variations or some combination of all the aforementioned effects.

\subsection{The Low-field Limit}
\label{subsec:lowfield}
We now analyze the data for low applied magnetic fields, where ``low''
specifies the field regime in which the SW model does not reproduce
the experimental behavior.  From this deviation, it is clear that more
complex magnetic configurations than a collective of parallel,
coherently rotating spins occur in this regime.

In literature, several non-trivial magnetic configurations are
suggested to play a role in magnetization reversal for core-free
systems such as the measured nanotubes. For the field parallel to the
tube axis there is the ``twisted bamboo'' state, where two vortex
states form at the ends of the tube with a domain parallel to the axis
between them \cite{wang_spin_2005, landeros_equilibrium_2009,
  chen_magnetization_2010, chen_magnetization_2011,
  buchter_reversal_2013}.  The magnetization reversal is thought to
take place by a propagating vortex or a transverse domain wall
\cite{landeros_reversal_2007, landeros_equilibrium_2009,
  bachmann_size_2009, landeros_domain_2010,
  albrecht_experimental_2011, streubel_rolled-up_2013}.  For a field
applied perpendicular to the nanotube axis, R\"{u}ffer et
al. \cite{ruffer_magnetic_2012} suggest the presence of an ``onion
state''. This configuration consists of two oppositely oriented
circumferential domains separated by two domain-walls, with the latter aligned parallel to the easy anisotropy axis and anti-parallel to each other.

The existence of such magnetic configurations has been predicted by
both analytical and numerical calculations
\cite{castano_metastable_2003, wang_spin_2005, topp_internal_2008,
  landeros_equilibrium_2009, landeros_domain_2010,
  streubel_equilibrium_2012, ruffer_magnetic_2012,
  buchter_reversal_2013, streubel_magnetic_2014}.  So far, however,
magnetic images of the configurations with sufficient spatial
resolution to clearly identify the states have not been possible.
Techniques that may produce images of sufficient resolution in the
future include scanning SQUID magnetometry, scanning diamond
nitrogen-vacancy (NV) center magnetometry
\cite{rondin_magnetometry_2014}, magnetic force microscopy (MFM)
\cite{castano_metastable_2003, li_flux_2001}, and X-ray circular
dichroism-photoelectron emission microscopy (XMCD-PEEM)
\cite{streubel_magnetically_2012,streubel_equilibrium_2012}.

Our DCM measurements cannot unambiguously determine the magnetic
configurations involved in the magnetization reversal.  Nevertheless,
guided by the SW model and micromagnetic simulations of the DCM
response, DCM yields important information about the progression of
magnetic configurations as a function of external field.  In
particular, given a particular progression of magnetic configurations
suggested by analytical or numerical calculations, DCM can be used to
support or rule out the hypothesis.

\subsubsection{Field Applied Parallel to Nanotube Axis}
\label{subsec:expParallel}
\begin{figure*}[t]
	\centering
	\includegraphics[width=2\columnwidth]{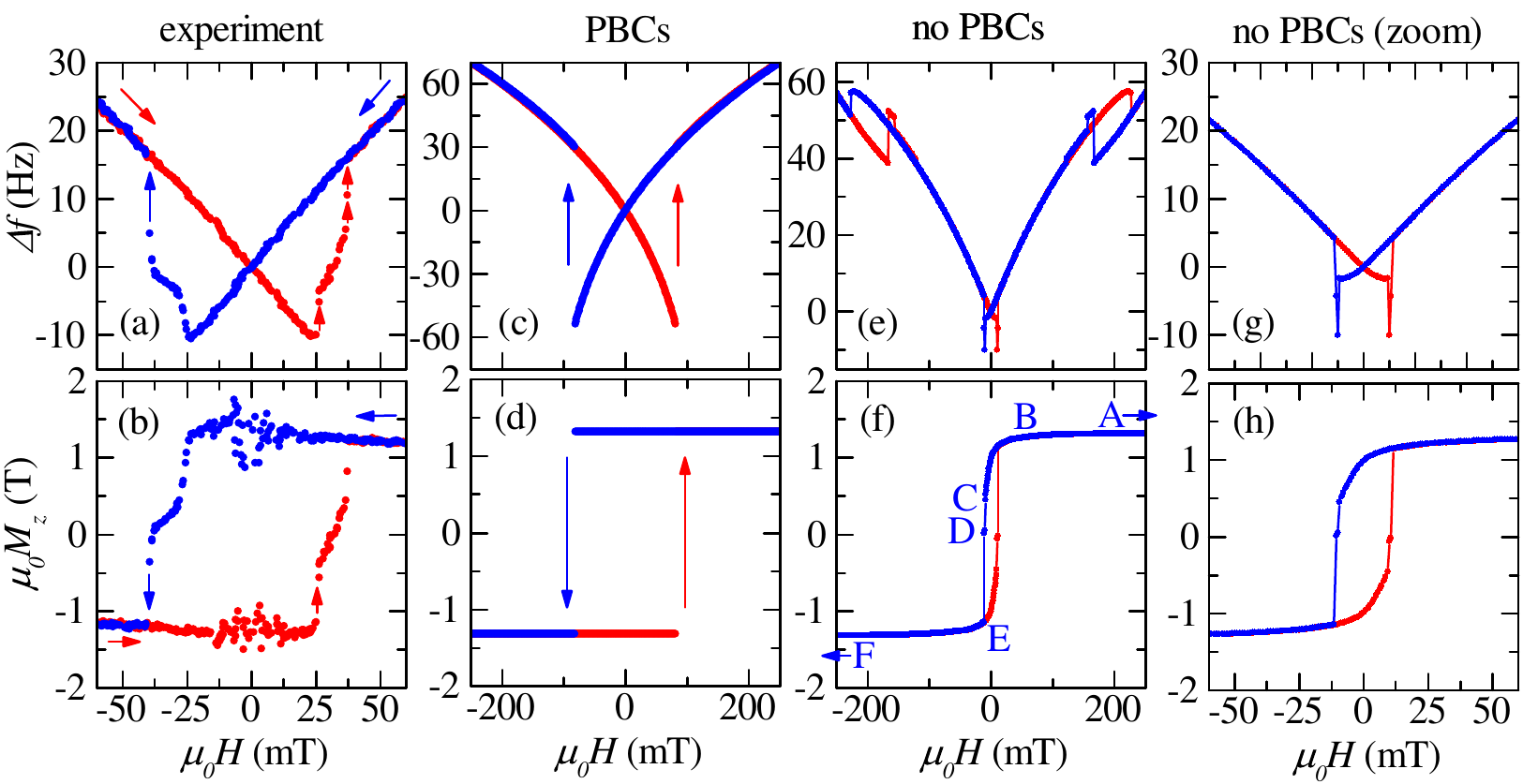}
	\caption{Frequency shift $\Delta f$ and magnetization $\mu_0
          M_z$, in the top and bottom rows respectively, vs.\ $\mu_0
          H$ for nanotube configuration~1, cf.\
          Fig.~\ref{fig:DCMsetup}.  Red (blue) curves are measurements
          for field sweeps in positive (negative) field direction. The
          first column shows DCM measurements, the second numerical
          simulations with PBCs, the third numerical simulations
          without PBCs, and the fourth a zoom of column three. }
	\label{fig:lowfield_C1}
\end{figure*}

\paragraph{DCM Measurements}
For configuration~1, we plot in Fig.~\ref{fig:lowfield_C1}(a) the same
data as in Fig.~\ref{fig:highfield}, focusing on the low magnetic
field range.
Here, as for high fields, the fit based on the SW model agrees well
with much of the measurement. The alignment of the easy axis with the
applied field direction combined with the strong shape anisotropy of
the nanotube keep all the magnetic moments parallel to each other and
aligned with $\mathbf{\hat{z}}$ for the majority of the magnetic field
range.  Despite this agreement, magnetization reversal takes place at
$\pm\SI{30}{\milli\tesla}$, rather than the
$\pm\SI{550}{\milli\tesla}$ predicted by the SW model. Also, a slight
asymmetry in the reversal fields and the switching behavior is likely
due to exchange coupling produced by a thin anti-ferromagnetic native
oxide on the nanotube surface \cite{buchter_magnetization_2015}.  Most
notably, however, the reversal takes place over three distinct stages
as shown in Fig.~\ref{fig:lowfield_C1}~(a): an initial step-like
feature, followed by a plateau near $\Delta f = 0$, ending with a
final irreversible magnetization switch.  Measurements of minor DCM
hysteresis loops show that both the initial and final steps are
irreversible.  In order to visualize the hysteresis curve more
clearly, we use (\ref{eq:MagVsDeltaf}) to extract the effective
macro-spin magnetization of the nanotube along $\mathbf{\hat{z}}$ from
the low-field frequency shift data.  The resulting low-field
hysteresis curve in Fig.~\ref{fig:lowfield_C1}~(b) shows a square loop
with a stable region with an effective magnetization of nearly zero in
the middle of both reversals.

Note that this DCM hysteresis curve differs from that observed for a
single Ni nanotubes by Weber et al.\ \cite{weber_cantilever_2012},
where magnetization reversal proceeded through a series of
statistically occurring steps attributed to multi-domain
states. Subsequent measurements by Buchter et
al. \cite{buchter_reversal_2013} indicated that the Ni nanotube
samples were separated in roughly \SI{0.5}{\micro\meter} long magnetic
segments due to the roughness of the film. In our case, the smoothness
of the CoFeB film and the absence of statistically occurring steps in
the magnetic hysteresis indicate that the CoFeB nanotubes are close to
ideal ferromagnetic tubes.
\begin{figure}[b]
	\centering
	\includegraphics[width=.9\columnwidth]{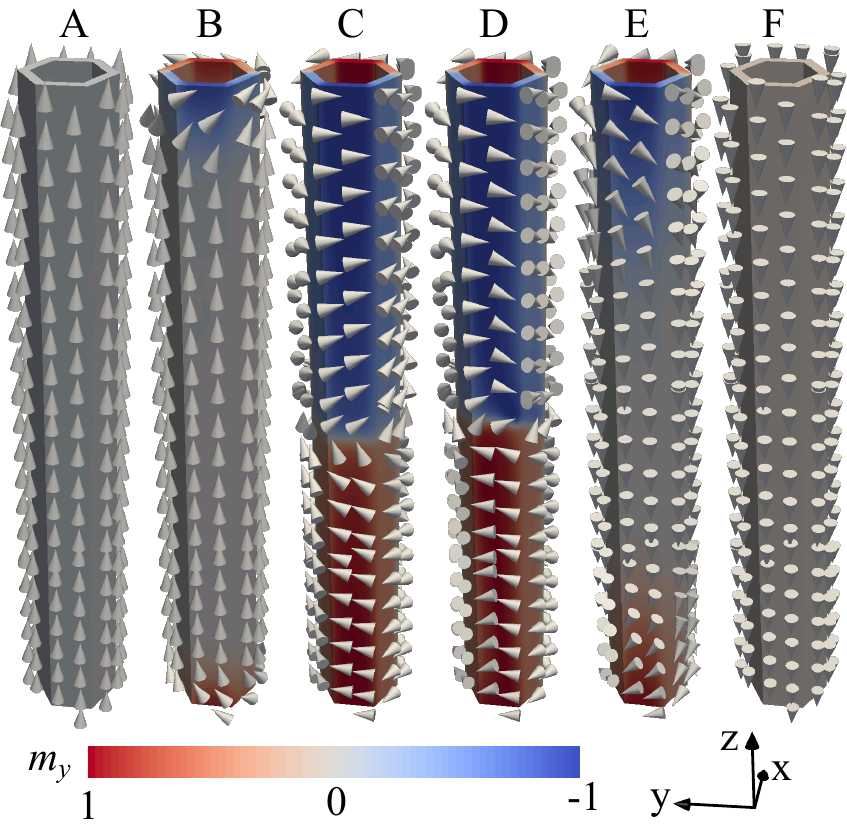}
	\caption{Visualization of the magnetization configurations
          calculated without PBCs for 6 different applied magnetic
          fields indicated in Fig.~\ref{fig:lowfield_C1}(f).  The
          applied field is swept from pointing in $+\mathbf{\hat{z}}$
          to $-\mathbf{\hat{z}}$ with $\mathbf{\hat{n}}$ pointing
          nearly along $\mathbf{\hat{z}}$ (configuration 1). }
	\label{fig:mag_config_C1}
\end{figure}

\paragraph{Numerical Simulations}
To establish a framework by which this hysteresis curve can be
understood, we analyze our micromagnetic simulations. In
Figs.~\ref{fig:lowfield_C1}~(c), (e), and (g), we plot the frequency
shift obtained from the simulation of a nanotube calculated with and
without PBCs.  In both cases, as discussed in
section~\ref{subsec:magSimulations}, we model a nanotube with a cross
section matching the measured tubes and a shorter length dictated by
computational limitations.  The simulated $\Delta f$ is then scaled up
by the ratio of the measured and simulated length of the nanotube
($\Delta f \propto V$) and $M_s$ is chosen for an optimal high-field
fit to the measurements, as described in
section~\ref{subsec:highfield}. Figs.~\ref{fig:lowfield_C1}~(d), (f),
and (h) show the magnetization along $\mathbf{\hat{z}}$ corresponding
to these simulations.

For simulations calculated with PBCs, $M_z$ performs a perfect rectangular hysteresis, and magnetization reversal takes place at $\pm\SI{100}{\milli\tesla}$, which is closer to the experimental value than obtained from the SW model. The magnetization distribution never shows any significant $M_x$ and $M_y$ components throughout the hysteresis curve and there is no trace of a plateau regime around $M_z = 0$. In fact, inspection of the simulated magnetization configurations reveal only the two axially saturated states throughout the hysteresis loop without the appearance of any non-uniform configurations.  Just as the SW approximation, this model, which ignores the effects of the nanotube ends, is inadequate for describing the observed low-field behavior.

The simulations calculated without PBCs show a magnetization reversal
characterized by the formation of two oppositely oriented vortices
nucleating at the two ends, just as in similar simulations carried out
by Buchter et al.\ \cite{buchter_reversal_2013} for Ni nanotubes.
Figs.~\ref{fig:mag_config_C1}A-F show the calculated magnetization
distribution, for various points in the hysteresis loop indicated in
Fig.~\ref{fig:lowfield_C1}~(f).  Coming from large, positive applied
field, following the dark blue data points, all spins are aligned
uniformly along the easy axis of the nanotube (A).  At around
$\SI{100}{\milli\tesla}$, oppositely oriented vortices nucleate at the
two ends due to the demagnetization effect of the end surfaces (B).
This is accompanied by an extremely small jump in the hysteresis curve
in Fig.~\ref{fig:lowfield_C1}~(f), which, however, has a visible
effect on $\Delta f$, cf.\ Fig.~\ref{fig:lowfield_C1}~(e).  The
vortices expand towards the center of the tube when approaching zero
field, separated by a N\'{e}el domain wall with a positive $M_z$
component (C).  This configuration is the so-called ``twisted bamboo''
state alluded to at the beginning of this section.  After crossing
zero field, this magnetic configuration persists, although $\Delta f <
0$ due to the sign change $H$.  At this point, $\Delta f$ approaches
zero as the system reaches a configuration with nearly zero net
magnetization along $\mathbf{\hat{z}}$ (D).  The magnetization in the central
part of the tube then undergoes an irreversible switch, corresponding
to a discontinuous retreat of the vortices to the ends and an
expansion of the central axial region (E).  Finally, the vortices
disappear at around $\SI{-200}{\milli\tesla}$ (F).  A clear signature
of this vortex mediated magnetization reversal seems to be the
discontinuity in $\Delta f$ when the vortices (dis)appear and the
rounding of $M_z$ close to the coercive field, which is also evident
in $\Delta f$, cf.\ Fig.~\ref{fig:lowfield_C1}~(e).  The simulations
show no step-like structure in the hysteresis near zero field.
\paragraph{Discussion}
The simulations calculated without PBCs capture the overall features
of the measured DCM and show the formation of a flux-closure
configuration near reversal. They do not, however, reproduce the
step-like feature in the measured reversal.  Also note that the
signatures of vortex nucleation in $\Delta f$, which are clearly
present in the simulations, are not observed in the measurement. This
discrepancy is likely due to the disproportionately large weight of
the nanotube ends in the simulation.  The simulations are based on a
\SI{1.5}{\micro\meter} tube, whose DCM response has been scaled up to
match the response of the greater than \SI{10}{\micro\meter} long
measured tube.  As a result, the fraction of the magnetic volume
occupied by the ends is disproportionately large in the simulated tube
in comparison with the measured tube. It is therefore possible that
the effect of the vortex nucleation on the measured $\Delta f$ is too
weak to clearly appear.

In the future, such discrepancies may be cleared up by increasing computational power and simulating tubes as long as those that are measured.  Conversely, improved sample preparation techniques may allow the measurement of tubes shorter than \SI{1.5}{\micro\meter}. For now, we are left to conclude that simulations showing reversal via flux-closure configurations most closely describe the measured reversal in a parallel applied field. Although we measure a step-like feature in the magnetization hysteresis reminiscent of a stable low-field flux-closure configuration, this feature is not reproduced by the simulations.

Nevertheless, we note that the step-like feature is similar to that
observed in magnetization curves of Co rings hosting a stable
low-field flux-closure configuration \cite{li_flux_2001,
  castano_magnetic_2004}. This resemblance suggests the possibility of
a stable low-field flux-closure configuration in the measured
nanotubes. At low fields, the single-domain configuration pointing
along the nanotube's long axis is dominated by its magnetostatic
self-energy. On the other hand, a flux-closure configuration, such as
a single vortex or the ``twisted bamboo'' configuration shown in
Fig.~\ref{fig:mag_config_C1}C and D, has reduced magnetostatic
self-energy and an increased exchange energy. For a magnetic nanotube,
however, the contribution of the vortex configuration to the exchange
energy is small: the absence of the magnetic core precludes the
formation of a central vortex, which, in nanowires for example,
provides a large contribution to the exchange energy.  As a result of
this reduced exchange contribution in nanotubes, a flux-closure
configuration can be the lowest energy state, and thus the stable
configuration at low fields.

We speculate that imperfections in the real sample -- not included in
the simulations -- could alter the hysteresis and stabilize a
low-field magnetic configuration.  Possible imperfections include the
non-ideal termination of the nanotube ends, an anti-ferromagnetic
native oxide layer, or magnetic pinning sites. However, while a
flux-closure configuration is among the possibilities explaining the
step-like feature in the hysteresis, other configurations with no net
magnetization along $\mathbf{\hat{z}}$ cannot be excluded, e.g.\
segments of opposing uni-axially aligned domains.
\subsubsection{Field Applied Perpendicular to Nanotube Axis}
\label{subsec:expPerpendicular}
\paragraph{DCM Measurements}
For configurations 2 and 3, the magnetization should execute the same
progression as a function of $H,$ since the external field is in both
cases applied perpendicular to the nanotube long axis.  In each
orientation, however, we probe the anisotropy in a different plane of
the nanotube leading to a different $\Delta f (H)$.  We focus
exclusively on an analysis of configuration~2, given the difficulty
disentangling the effects of sample misalignment from the anisotropy
in the nanotube cross-section as discussed in
section~\ref{subsec:highfield}.
\begin{figure*}[t]
	\centering
	\includegraphics[width=2\columnwidth]{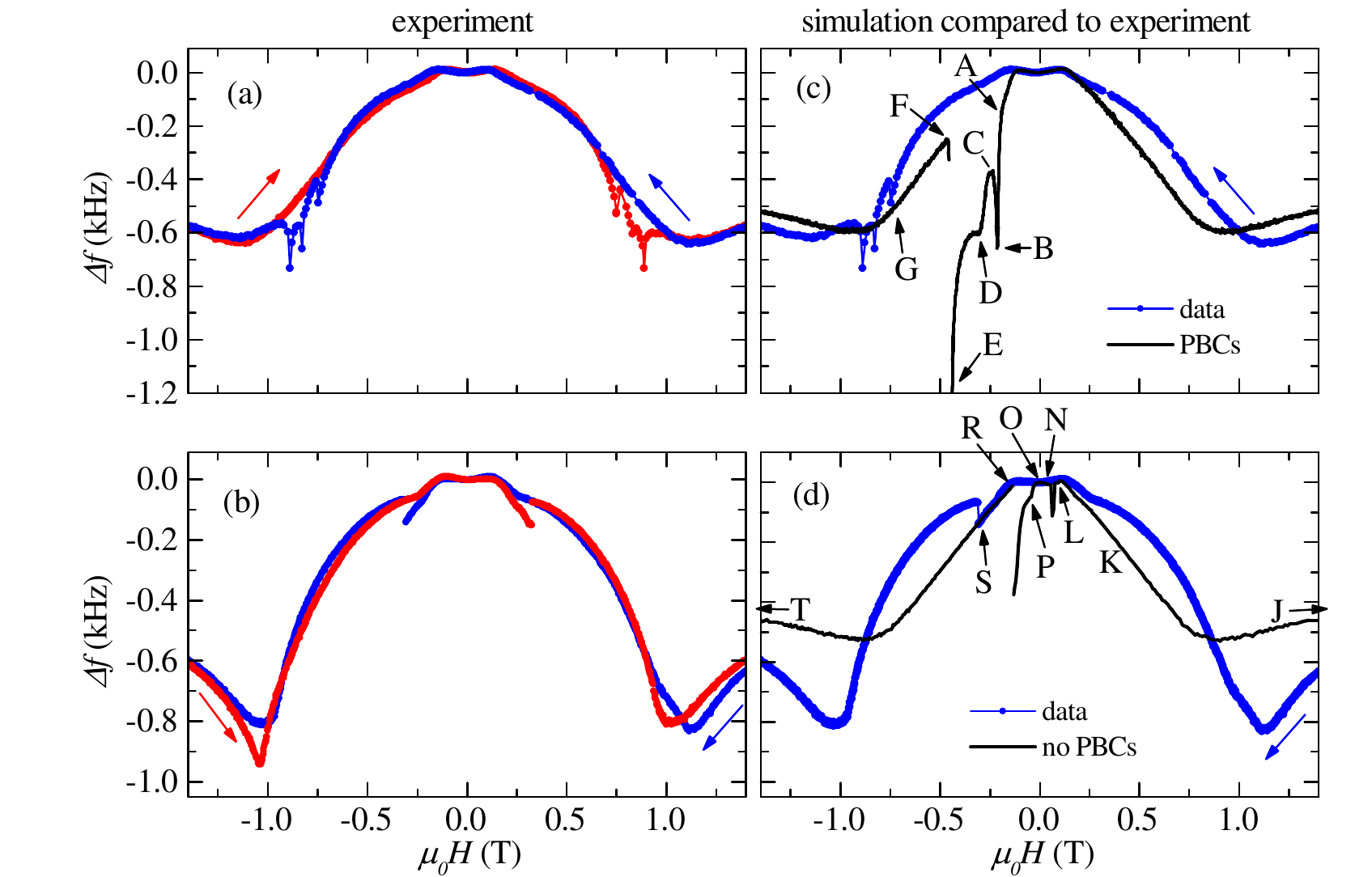}\\
	\caption{Frequency shift $\Delta f$ vs $\mu_0 H$ for nanotube
          configuration~2, cf.\ Fig.~\ref{fig:DCMsetup}.  Red (blue)
          curves are measurements for field sweeps in positive
          (negative) field direction.  Two different types of
          hysteresis loops are observed to occur, depicted in (a) and
          (b). (c) compares the measured data from (a), with
          simulations calculated with PBCs in black.  (d) compares the
          measured data from (b) with simulations calculated without
          PBCs in black.}
	\label{fig:lowfield_C2}
\end{figure*}

For $\mu_0 |H| \gtrsim \SI{1}{\tesla}$, the fit based on the SW model
describes the data well, as discussed in
section~\ref{subsec:highfield}.  This agreement shows the presence
of a single-domain configuration with all magnetic moments pointing
along the field direction, cf.\ the last column of charts in
Fig.~\ref{fig:greatgraph}.  In the SW model, the two minima of $\Delta
f$ mark the field magnitude, below which $\mathbf{M}$ starts rotating
towards the long axis of the nanotube. This minimum occurs at $\mu_0
|H| \approx \SI{1}{\tesla}$ for the measurement.  In
Figs.~\ref{fig:lowfield_C2}(a) and (b) we show the data for this lower
field regime, where the coherent SW reversal is likely to be
superceded by a more complex behavior.  In this regime a double hump
feature is observed in $\Delta f (H)$, which shows hysteretic behavior
and can include either a discontinuous jump close to one of the humps
(see Fig.~\ref{fig:lowfield_C2}(b)) or three discontinuous spikes
between $\SI{0.5}{\tesla}$ and $\SI{1.0}{\tesla}$ (see
Fig.~\ref{fig:lowfield_C2}(a)).  These two different types of $\Delta
f(H)$ curves occur statistically and, though qualitatively similar,
include consistent differences for $\mu_0|H| \lesssim \SI{1}{\tesla}$.

\paragraph{Numerical Simulations}
In Figs.~\ref{fig:lowfield_C2}(c) and (d) we compare the measured
curves for decreasing magnetic field (blue) with the results from the
finite element simulations (black), and find an overall agreement
between the curves, where all major features are reproduced.  The
low-field double hump feature is well-reproduced in both the
simulations with and without PBCs, especially in comparison with the
poorly matching low-field $\Delta f (H)$ predicted by the SW model,
shown in Fig.~\ref{fig:highfield}(b).  We can therefore identify this
low-field feature as an effect of the hexagonality of the nanotube's
cross-section, which is absent from the SW model and present in both
numerical simulations.

Interestingly, the discontinuities observed for the progression shown
in Fig.~\ref{fig:lowfield_C2}(b) and for the simulated nanotube
without PBCs are similar, while the ones observed for the progression
shown in Fig.~\ref{fig:lowfield_C2}(a) resemble those seen in the
simulations with PBCs.  The latter shows three strong spikes after
passing zero field, when coming from positive field, that are also
observable in experiment, although weaker.  As discussed in the
following, we find that all these discontinuities arise due to an
interplay between the imperfect alignment of the nanotubes and their
hexagonal cross-section.
 
\begin{figure}[t]
	\centering
  \includegraphics[width=.99\columnwidth]{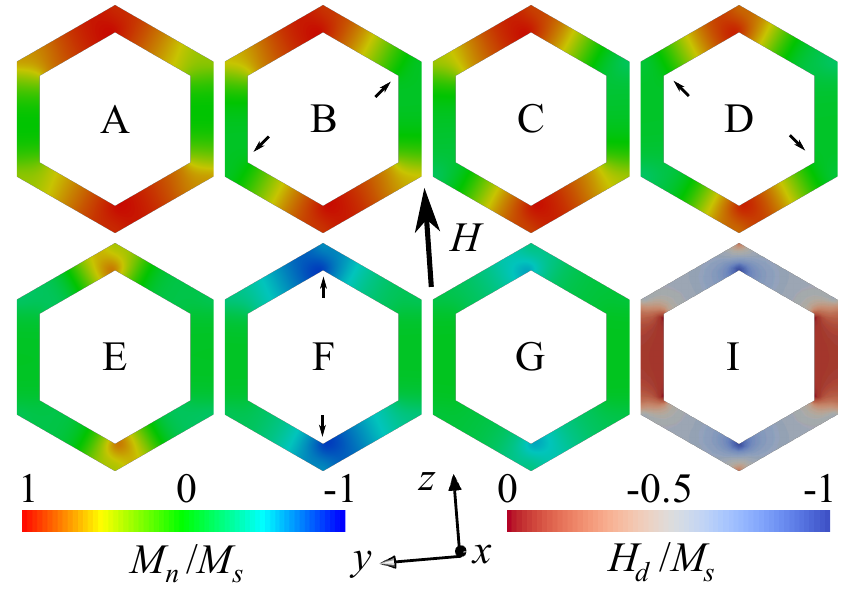}
  \caption{Simulated magnetization $M_{\hat{n}}/M_s$ with PBCs. Images A-G depict the $M_{\hat{n}}$ component of magnetization in the nanotube cross-section for the corresponding field values indicated in Fig.~\ref{fig:lowfield_C2}(d).  Image I shows the local demagnetizing field in $\mathbf{\hat{z}}$ direction $H_d/M_s$ for an applied field of $H=1.15 M_s$. In all cases the applied field $\mathbf{H}$ is slightly misaligned from being perpendicular to the nanotube long axis, as shown by the arrow. $\theta_n = 86^{\circ}$ and $\phi_n = 5^{\circ}$ as extracted from the high-field SW fits for data obtained in configuration 2, c.f.\ Table~\ref{tab:fitparameters}.}
	\label{fig:hexdemag}
\end{figure}

As shown in Fig.~\ref{fig:DCMsetup}(d), in both configurations 2 and
3, the hexagon is aligned such that its two lateral facets are
parallel -- or in the usual case of slight misalignment, nearly
parallel -- to the applied field.  Within such a hexagonal
cross-section, the demagnetization field opposing the external field
varies strongly as a function of position, cf.\
Fig.~\ref{fig:hexdemag}I.  As a result, sweeping an external field
applied perpendicular to the long axis of the nanotube down from large
values, moments in the top and bottom vertices of the hexagon will
start rotating toward the long axis before moments in other locations.
Moments in the four top and bottom facets will next begin rotating,
while moments in the two side facets will rotate at the smallest
fields.  As the field is increased from zero again, this staggered
rotating occurs in reverse order.

Furthermore, given the inevitable misalignment of $\theta_n$, the
magnetic moments prefer to rotate towards one of the two easy axis
directions, when coming from large fields: e.g.\ towards
$\mathbf{\hat{n}}$ rather than $-\mathbf{\hat{n}}$ coming from
positive $H$ in Fig.~\ref{fig:NT_structure}. This preferred direction
depends on the direction from which the field is swept, i.e.\
misalignment from perfect perpendicularity with $\mathbf{H}$
introduces hysteresis. Therefore, at zero field, the magnetization
tends to point along the easy axis direction preferred by the sample's
magnetic history. Upon reversal of the applied field direction, the
other easy axis direction becomes energetically favorable, ultimately
resulting in a discontinuous reversal.  This reversal along the easy
axis produces the discontinuities observed in both DCM measurements
and simulations.

With this insight, let us now follow the simulated $\Delta f(H)$-curve
calculated with PBCs in Fig.~\ref{fig:lowfield_C2}(c), which most
closely matches the behavior measured in
Fig.~\ref{fig:lowfield_C2}(a).  As the field is swept down from
positive fields near zero, the magnetization tends to point along its
preferred easy axis direction $\mathbf{\hat{n}}$. Reducing the field
past zero, the moments in the two side facets begin to rotate towards
the field direction, as can be seen in the local distribution of $M_n$
in Fig.~\ref{fig:hexdemag}A ($M_n\approx 0$ in the side facets), where
$M_n$ is the component of $\mathbf{M}$ pointing in direction of the
easy axis $\mathbf{\hat{n}}$.  As with the other configurations, this
configuration corresponds to the field value indicated by the
corresponding letter in Fig.~\ref{fig:lowfield_C2}(c).  Next, as
indicated by two black arrows in Fig.~\ref{fig:hexdemag}B, the
magnetic moments in two of the side vertices rotate, leading to a
strong negative spike in $\Delta f$.  $\Delta f$ recovers for a small
field range, shown as C, until the moments in the other two side
vertices rotate, leading to a somewhat smaller spike, D. From here,
$\Delta f$ further decreases and the moments in the top and bottom
facets rotate, as shown in E.  A last jump occurs as the moments in
the top and bottom vertices reverse their $M_n$, depicted in
F. Finally, G shows the distribution with which $M_n$ decreases as the
magnetic moments complete their rotation towards $-\mathbf{\hat{z}}$.
Although appearing in a slightly different field range and with
smaller magnitude, the three spikes in the $\Delta f(H)$-curve
simulated with PBCs are also clearly observable in the experiment,
cf.\ Fig.~\ref{fig:lowfield_C2}(a) and (c).  This effect is likely be
due to the imperfect hexagonality of the measured nanotube.  Note that
for a perfect alignment ($\theta_n=90^{\circ}$), one would expect the
discontinuities shown at B and D to merge, since by symmetry the
vertices shown in Fig.~\ref{fig:hexdemag}B and D must reverse at the
same applied field.

\begin{figure}[t]
	\centering
	\includegraphics[width=1\columnwidth]{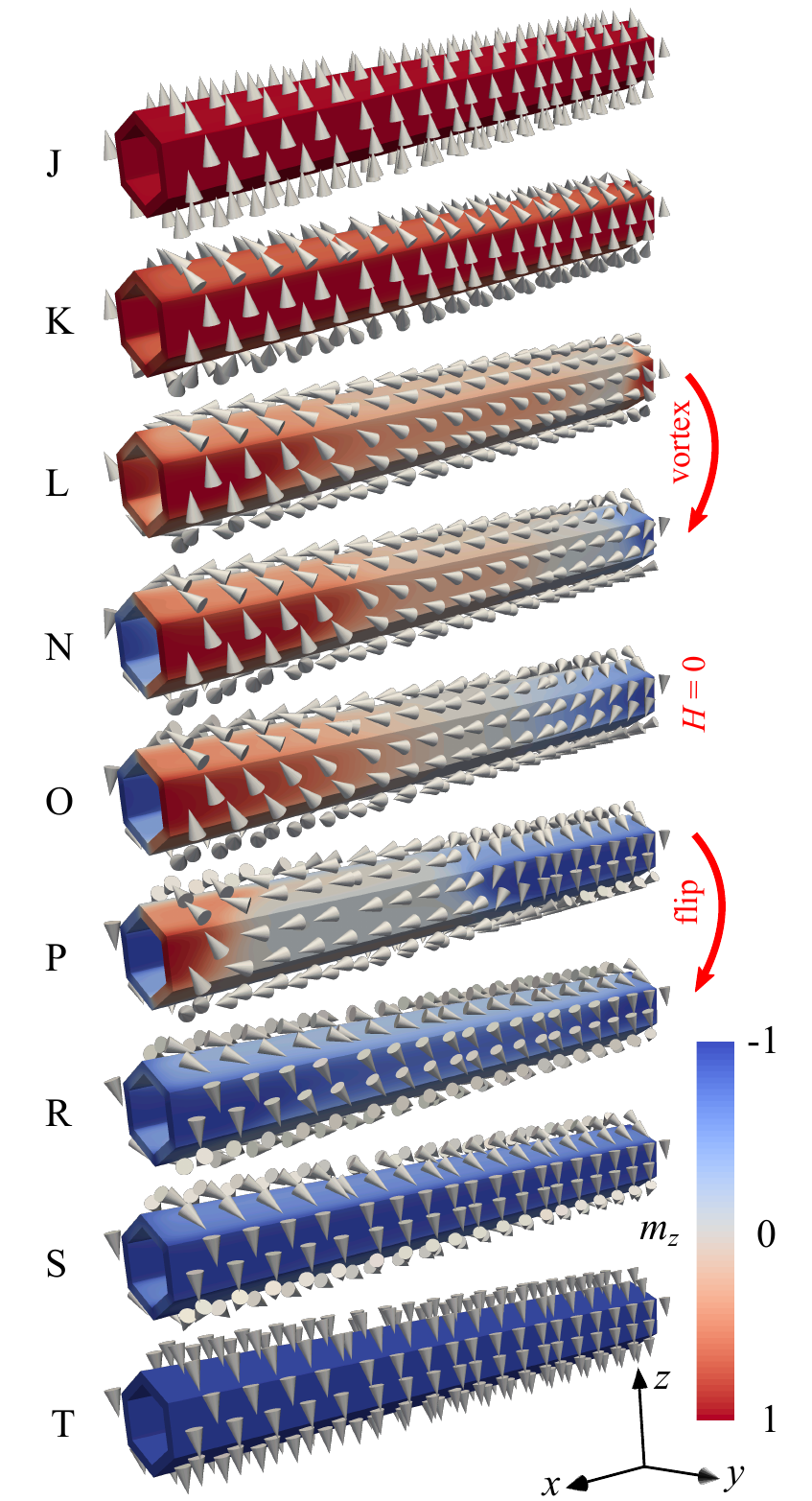}\\
        \caption{Visualization of the magnetization configurations
          calculated without PBCs for 9 different applied magnetic
          fields indicated in Fig.~\ref{fig:lowfield_C2}(d).  The
          applied field is swept from pointing in $+\mathbf{\hat{z}}$ to
          $-\mathbf{\hat{z}}$ with $\mathbf{\hat{n}}$ pointing nearly along $\mathbf{\hat{x}}$
          (configuration 2 or 3).}
	\label{fig:progression_C2}
\end{figure}

In contrast to the the $\Delta f (H)$-curve simulated with PBCs, the
curve simulated without PBCs has a single discontinuity, which occurs
for a smaller applied field than the three spikes, as shown in
Fig.~\ref{fig:lowfield_C2}(d).  A second, even smaller discontinuity
is seen just before crossing zero field.  To shed light on the
calculated progression of magnetic configurations, we show in
Fig.~\ref{fig:progression_C2} images of the magnetic configurations
corresponding to the points indicated in
Fig.~\ref{fig:lowfield_C2}(d).

Figure~\ref{fig:progression_C2}J shows the tube magnetization saturated
along $\mathbf{\hat{z}}$ at large positive field. As the field is reduced
across K and then L, we observe the staggered tilting of magnetic
moments imposed by the differences in demagnetization field within the
hexagonal cross-section of the tube (Fig.~\ref{fig:hexdemag}I).  The
behavior of the tube ends differs from that of the central part.
Between L and N, vortices nucleate at the ends, producing the
corresponding discontinuity for small positive field in
Fig.~\ref{fig:lowfield_C2}(d).  At zero field, O, the vortices have
maximally expanded along the tube, separated by a region of magnetic
moments pointing along the preferred easy axis direction.  This
configuration is essentially the ``twisted bamboo'' state consisting
of two opposing vortex configurations separated by a domain pointing
along the tube axis.  This configuration also appeared near zero field
in the simulations calculated without PBCs for configuration 1, i.e.\
Fig.~\ref{fig:mag_config_C1}C and D. The only difference is that, in
the present case, the domain parallel to the tube axis is more extended. In
P, the moments continue to rotate along the negative applied field.
Now, however, since the preferred easy axis direction has reversed,
the moments in the parallel domain find themselves pointing against
the most energetically favorable direction.  The vortices seem to
stabilize the overall magnetic configuration, and, as a consequence, a
single discontinuous change occurs between P and R, resulting in a
nearly uniform configuration without vortices, pointing mostly between
the negative applied field and the preferred easy axis.  S and T then
show the final staggered rotation of the magnetic moments along the
applied field.  The simulated $\Delta f(H)$-curve corresponding to
this progression strongly resembles the reversal measured by DCM shown
in Fig.~\ref{fig:lowfield_C2}(b).  The absence of the discontinuity
due to the vortex nucleation in the measurement is likely due to the
disproportionately large weight of the nanotube ends in the
simulation, as already discussed in Section~\ref{subsec:expParallel}.

\paragraph{Discussion}
The two types of DCM reversals shown in Fig.~\ref{fig:lowfield_C2}(a)
and (b) occur statistically and resemble simulations carried out with
PBCs and without, respectively.  Therefore, we hypothesize that the
reversal in the perpendicular field geometry takes either the form of
a staggered rotation without the formation of vortices
(Fig.~\ref{fig:lowfield_C2}(a)) or a reversal in which vortices
nucleate and a low-field ``twisted bamboo'' state is traversed
(Fig.~\ref{fig:lowfield_C2}(b)) as depicted in
Fig.~\ref{fig:progression_C2}.
%
%
%
\section{Conclusion}
\label{sec:summary}
To conclude, we investigate magnetic states of individual CoFeB
nanotubes in an applied magnetic field for different tube-to-field
orientations using DCM.  Single nanotubes are attached to the end of
an ultra-soft cantilever and the shift of the cantilever's resonance
frequency due to the dynamic magnetic torque is detected.  We
introduce an analytical model for DCM of an idealized SW magnet, in
which all moments act in unison.  Applying this model to the
magnetometry data, we are able to describe behavior at high external
fields and extract a saturation magnetization $\mu_0 M_s = 1.3 \pm
\SI{0.1}{\tesla}$ for the CoFeB nanotubes.

In order to construct a more realistic model and to describe the
behavior of the nanotubes at low field, we develop a numerical
micromagnetic framework for calculating DCM frequency shifts and
implement it using \textit{Nmag}.  These numerical simulations show
that hysteresis loops measured by DCM in both parallel and -- in some
cases -- perpendicular applied field resemble what is expected for a
reversal sequence nucleated by vortices at the tube ends.  Such
reversals include a ``twisted bamboo'' flux-closure configuration at
low fields.  Although the measurements in parallel field show the
signature of a stable configuration near zero-field, corresponding
simulations do not confirm the behavior.  This discrepancy may be due
to the stabilizing effect of imperfections in real samples on magnetic
configurations at low field.

Future work to confirm the presence and spatial character of these
configurations should focus on non-invasive imaging of the nanotubes'
magnetization.  Potentially applicable techniques include scanning
SQUID magnetometry, diamond NV center magnetometry
\cite{rondin_magnetometry_2014}, MFM \cite{castano_metastable_2003,
  li_flux_2001}, XMCD-PEEM
\cite{streubel_magnetically_2012,streubel_equilibrium_2012}.  Finally,
we note that our numerical method for calculating DCM response is not
specific to ferromagnetic nanotubes and is therefore applicable to a
broad range of nanomagnetic samples.  The framework may allow us to
accurately interpret the DCM response of other nanometer-scale
ferromagnets or even of anti-ferromagnets and helimagnets.
%
%
\begin{acknowledgments} We thank G\"{o}zde T\"{u}t\"{u}nc\"{u}oglu
  and Federico Matteini for providing the GaAs NW used as templates
  for the CoFeB nanotubes. We gratefully acknowledge financial support
  from the Canton Aargau, the Swiss Nanoscience Institute (SNI), the
  Swiss National Science Foundation (SNSF) under Grant
  No. 200020-159893, the NCCR Quantum Science and Technology (QSIT),
  and the Deutsche Forschungsgemeinschaft via project GR1640/5-2
  (Priority Program SPP 1538).
\end{acknowledgments}

\end{document}